\documentclass[prd,preprintnumbers,amsmath,amssymb,superscriptaddress]{revtex4}  %twocolumn,
\usepackage{graphicx}% Include figure files  
\usepackage{epsfig}% Include figure files
\usepackage{rotating}% Include figure files
\usepackage{dcolumn}% Align table columns on decimal point
\usepackage{bm}% bold math
\newcommand{\beq}{\begin{eqnarray}} 
\newcommand{\eeq}{\end{eqnarray}} 

\begin{document}

\preprint{LPT-ORSAY-12-30}

%% \linenumbers 

\title{Higgs boson production via vector-like top-partner decays: \\
diphoton or multilepton plus multijets channels at the LHC}

\author{A.~Azatov}
\affiliation{Dipartimento di Fisica, Universit\`a di Roma ``La Sapienza'', INFN Sezione, I-00185 Roma, Italy}
%\email{}
%%%%%%%%%%%%%%%%%%%%%%%%%%%%
\author{O.~Bondu}
\affiliation{Universit\'e de Lyon,
Universit\'e Claude Bernard Lyon 1, CNRS–IN2P3, Institut de Physique
Nucl\'eaire de Lyon, F-69622 Villeurbanne Cedex, France}
%%%%%%%%%%%%%%%%%%%%%%%%%%%%
\author{A.~Falkowski}
\affiliation{Laboratoire de Physique Th\'eorique, B\^at. 210, CNRS,
Universit\'e Paris-sud 11, F-91405 Orsay Cedex, France}
%%%%%%%%%%%%%%%%%%%%%%%%%%%%
\author{M.~Felcini}
\affiliation{Instituto de F\'isica de Cantabria (IFCA),
CSIC-Universidad de Cantabria, E-39005 Santander, Spain}
%%%%%%%%%%%%%%%%%%%%%%%%%%%%
\author{S.~Gascon-Shotkin}
\affiliation{Universit\'e de Lyon,
Universit\'e Claude Bernard Lyon 1, CNRS–IN2P3, Institut de Physique
Nucl\'eaire de Lyon, F-69622 Villeurbanne Cedex, France}
%%%%%%%%%%%%%%%%%%%%%%%%%%%%
\author{D.~K.~Ghosh}
\affiliation{Department of Theoretical Physics, Indian Association
for the Cultivation of Science Kolkata 700 032, India}
%%%%%%%%%%%%%%%%%%%%%%%%%%%%
\author{G.~Moreau}
\affiliation{Laboratoire de Physique Th\'eorique, B\^at. 210, CNRS,
Universit\'e Paris-sud 11, F-91405 Orsay Cedex, France}
%%%%%%%%%%%%%%%%%%%%%%%%%%%%
\author{A.~Y.~Rodr\'iguez-Marrero}
\affiliation{Instituto de F\'isica de Cantabria (IFCA),
CSIC-Universidad de Cantabria, E-39005 Santander, Spain}
%%%%%%%%%%%%%%%%%%%%%%%%%%%%
\author{S.~Sekmen}
\affiliation{CERN Physics Department, CH-1211 Geneva 23, Switzerland.}
%%%%%%%%%%%%%%%%%%%%%%%%%%%%

\begin{abstract} 
We first build a minimal model of vector-like quarks where the dominant Higgs boson production process at LHC -- the gluon fusion -- can be significantly suppressed,
being motivated by the recent stringent constraints from the search for direct Higgs production over a wide Higgs mass range.
Within this model, compatible with the present experimental constraints on direct Higgs searches, 
we demonstrate that the Higgs ($h$) production via a heavy vector-like top-partner ($t_2$) decay, 
$pp \to t_2 \bar t_2$, $t_2\to t h$,  allows to discover a Higgs boson at the LHC and measure its mass, 
through the decay channels $h\to \gamma\gamma$ or  $h\to ZZ$.
We also comment on the recent hint in LHC data
from a possible $\sim 125$~GeV Higgs scalar, in the presence of heavy vector-like top
quarks.
\end{abstract} 
%\pacs{} 

\maketitle

%%% \begin{linenumbers}  

\section{INTRODUCTION}

One of the primary goals  of the  Large Hadron Collider (LHC) is  the direct search for
the cornerstone of the Standard Model (SM),
namely the Higgs boson, or for any signal from alternative
Electro-Weak Symmetry Breaking (EWSB) mechanisms. The SM is
probably not the ultimate model of nature. 
It is clear that new channels for  Higgs production, 
that can arise in extensions of the SM, would have profound
impact on the discovery of a Higgs boson, while providing  insight in the physics beyond the SM.
An attractive possibility is the Higgs production in decays of
additional heavy colored particles that can be copiously
pair produced
at the LHC via strong interactions.

Within well-motivated theories beyond the SM, there are some
candidates for such new heavy colored states,
extra quarks with vector-like couplings, whose existence is
predicted by most of the alternatives to supersymmetry.
In this context, to maintain a naturally
light Higgs boson, divergent quantum corrections from loops of
the top quark are often canceled by top-partner
contributions~\cite{ArkaniHamed:2001nc,ArkaniHamed:2002qx,ArkaniHamed:2002qy}.
Let us describe important examples here. 
In the so-called little Higgs scenarios, the vector-like quarks
arise as partners of the SM fields being promoted to larger
multiplets. 
In the composite
Higgs~\cite{Contino:2003ve,Agashe:2004rs,Agashe:2005dk,Contino:2006qr,Burdman:2007sx}
and composite
top~\cite{Contino:2003ve,Agashe:2004rs,Agashe:2005dk,Contino:2006qr,Burdman:2007sx,Hill:1991at}
models, the vector-like quarks
are excited resonances of the bounded states constituting the SM
particles.
In the extra-dimensional models with (SM) quarks in the bulk,
the vector-like quarks are prevalent as Kaluza-Klein (KK)
excitations of those
bulk fields~\cite{Cheng:1999bg} like in the Gauge-Higgs
unification mechanism (see e.g.
Ref.~\cite{Carena:2006bn,Carena:2007ua})
or in the Randall-Sundrum (RS)
scenario~\cite{Randall:1999ee,Gogberashvili:1998vx,Gherghetta:2000qt}
-- where some of those KK excitations, the so-called custodians,
can be as light as a few hundred's of
GeV~\cite{Agashe:2003zs,Bouchart:2008vp,Djouadi:2006rk,Djouadi:2007eg,Agashe:2006at,Bouchart:2009vq}.
Another example is a gauge coupling unification theory where
vector-like quarks
are embedded into the simplest ${\rm SU(5)}$
representations~\cite{Kilic:2010fs}.

Vector-like quarks with same electric charge as the up-type quarks are often called top-partners (noted
$t'$)
as these new heavy states mix in general predominantly with the
top quark -- due to the large top mass and to the related
feature that the top
quark is in general more intimately connected to ultraviolet
physics, like e.g. in composite Higgs models.
A $t'$ can also be called a top-partner in the sense that it is
contained in the same group representation
as the top quark with respect to symmetries, like the
approximate global symmetry of the little Higgs
models~\cite{ArkaniHamed:2001nc,ArkaniHamed:2002qx,ArkaniHamed:2002qy},
the gauge unification symmetry~\cite{Kilic:2010fs} or
the custodial symmetry of RS versions with bulk
matter~\cite{Agashe:2003zs,Bouchart:2008vp,Djouadi:2006rk,Djouadi:2007eg,Agashe:2006at,Bouchart:2009vq}
(explaining the SM fermion mass
hierarchies~\cite{Huber:2000ie,Huber:2001ug,Huber:2002gp,Huber:2003sf,Chang:2005ya,Moreau:2006np,Moreau:2005kz,Agashe:2004cp,Agashe:2004ay,Agashe:2006wa,Agashe:2006iy,Grossman:1999ra,Appelquist:2002ft,Gherghetta:2003he,Moreau:2004qe,delAguila:2008iz,Raidal:2008jk}).

At this level, one must mention that
the phenomenology of the search for direct production of
vector-like quarks at the LHC has been studied from
a model-independent point of view in
Ref.~\cite{Cacciapaglia:2010vn,Cacciapaglia:2011fx,Gopalakrishna:2011ef,AguilarSaavedra:2005pv,AguilarSaavedra:2009es,Ivanov:2011zz}
but also in specific frameworks such as the little Higgs models
(versions sufficiently safe from EW precision
constraints)~\cite{Burdman:2002ns,Han:2003wu,Perelstein:2003wd,Cheng:2005as,Cacciapaglia:2009cu}
or the composite Higgs hypothesis~\cite{Vignaroli:2011ik,Contino:2008hi}
and the dual RS context~\cite{Dennis:2007tv,Agashe:2004bm,Brooijmans:2010tn,Carena:2006bn,Bini:2011zb,Barcelo:2011vk,Barcelo:2011wu}.
These past searches focus generally on the discovery of the
vector-like quarks, rather than using these extra quarks
to enhance the discovery and identification potential for other
unknown particles such as Higgs scalars.

In relation to Higgs detection, there exist studies utilizing the possible Higgs production through vector-like quark decays,
as described in the following. Indeed, it is well known since some time~\cite{delAguila:1989rq,delAguila:1989ba}
that vector-like quark production could be a copious source of Higgs bosons (a possible Higgs factory). 
\\ Relatively light Higgs bosons produced from the decay of top-partners 
can be highly boosted and good candidates for analyses based
on jet substructure.  This method
has been applied~\cite{Kribs:2010ii} for a $\sim$130 GeV Higgs
decaying to $b\bar b$ at the $14$~TeV LHC
to improve the Higgs identification capability and reduce the background. 
In the simple model considered there, the $t'$ is a singlet under the ${\rm SU(2)_L}$ gauge group,
which determines the $t'$ couplings and its tree level  decays into 
the Higgs boson and the two EW gauge bosons $t'\to th$, $t'\to tZ$, $t'\to bW$.
\\ The top-partner can also be singly produced which leads to
different final states as compared to the pair production; because of the
phase space suppression, the single production becomes competitive with the pair
production at a high $t'$ mass, depending upon the considered model
(since the involved $t'$ couplings to $h,Z^0,W^\pm$ are fixed by
the $t'$ quantum numbers)~\footnote{The single production
becomes dominant
typically around $m_{t'}\sim 700$~GeV as in the scenario with a
singlet $t'$~\cite{AguilarSaavedra:2005pv} or in little Higgs
models~\cite{Azuelos:2004dm}.}.
The reconstruction of the Higgs boson produced in the $t'$
decay, itself singly produced at the $14$~TeV LHC, was studied
in Ref.~\cite{Azuelos:2004dm}
assuming the Higgs mass known (to be $120$~GeV) and focusing on
the channel $h\to b\bar b$ -- with the
combinatorial background only. 
This was performed for a singlet $t'$ in the ``Littlest Higgs''
model with the asymptotic branching ratio values of the high
$m_{t'}$ regime: $B_{t' \to t h}=25\%$, $B_{t' \to t Z}=25\%$,
$B_{t' \to b W} =50\%$ (from the EW equivalence theorem).
\\ Similarly, a vector-like colored $b'$ state produced at the
$14$~TeV LHC can act as a Higgs factory thanks to its decay
$b'\to bh$.
It was shown~\cite{Kilic:2010fs} that a Higgs mass
reconstruction can be obtained with a limited accuracy,
concentrating on the decay $h\to WW$
($W\to l\nu$) for $m_h=200$~GeV and assuming the $m_{b'}$ value
to be deduced from a preliminary analysis based on the more
appropriate
channel $b'\to bZ$. Theoretically, the $b'$ was originating from
the upper component of a ${\rm SU(2)_L}$ doublet so there was no
significant channel $b'\to t W$.   
\\ Higgs mass reconstructions via $t'$ and $b'$ decays were also
studied for the $14$~TeV LHC, based on a light Higgs decaying to
$b\bar b$
in the basic models with a unique extra $t'$ and/or a unique
extra $b'$~\cite{AguilarSaavedra:2009es}.

In the present paper, we use the pair production and decay of a
vector-like top to develop new search strategies for Higgs boson 
discovery and mass measurements in  the $h\to\gamma \gamma$ (diphoton) and  $h\to ZZ$ channels.
We consider $t'$ masses up to $\sim 800$~GeV, so that the $t'$ single productions (involving a model-dependent coupling)
are generally subleading compared to the $t'$ pair production not yet significantly suppressed by phase space factors~\footnote{Note
that adding the contribution of the $t'$ single production would increase the Higgs production rate.}.  
The original theoretical and illustrative model considered here, including two
top-partners, is constructed to allow interesting
interpretations correlating the
indirect (via vector-like top decay) and direct Higgs production searches at the
LHC, as described in the following.
A few characteristic parameter sets -- with vector-like top mass in the range between $\sim$~400 and 800~GeV
-- are chosen as benchmark points avoiding too large $t'$ contributions to the Higgs rates 
(constrained by present LHC data) and simultaneously allowing for significant branching fraction 
values ($\gtrsim 10\%$) of the vector-like top decay to the Higgs boson.  
\\ Assuming the presence at low-energy scales only of extra
vector-like quark multiplets containing some $t'$, we have
elaborated a minimal model allowing to strongly suppress the Higgs production via gluon fusion, as compared to the SM. 
In this simple but non-trivial model, the $gg\rightarrow h$ cross section suppression factor possibly reaches values 
below $10^{-1}$ at hadron colliders; this is to be put in contrast with 
the $t'$ representations taken usually in the RS scenario 
\cite{Bouchart:2009vq,Djouadi:2007fm,Djouadi:2009nb,Djouadi:2011aj,Azatov:2010pf}~\footnote{$t'$
representations (to which SM fields are promoted)
with analog rate suppression effects can arise with the ${\rm
O(3)}$ subgroup
\cite{Agashe:2006at} implementable in the composite Higgs
model~\cite{Falkowski:2007hz,Espinosa:2010vn}
and RS scenario~\cite{Ledroit:2007ik,Casagrande:2010si} which
can reach strong suppressions,
respectively of $\sigma_{\rm gg\to h}/\sigma_{\rm gg\to h}^{\rm
SM}\sim 35\%$ and $\sim 10\%$.}
and with minimal supersymmetric theories for which such a
suppression is not possible to obtain (see respectively
Ref.~\cite{Djouadi:2007fm,Bouchart:2009vq} and
Ref.~\cite{Djouadi:1998az}). The chiral case of a fourth quark
generation can even only increase
considerably the gluon fusion rate.
\\ The illustrative minimal $t'$ model suggested here is interesting in the
sense that it can easily lead to the following interpretations:
for example, a $255$~GeV Higgs is excluded in the SM by the present LHC
results~\cite{CMSweb,ATLASweb} but can still
exist in the above minimal SM extension with $t'$ where the
reduced Higgs production cross section can be below the LHC
upper limits.
In other words, the Higgs boson would really be light but not
detectable with the present luminosity/energy, via conventional
channels.
A channel that could then allow the Higgs discovery would be 
through the $t'$ pair production and decays, as illustrated in
this paper.
Another possibility is that the slight excess of events observed in data for a  Higgs
mass hypothesis of  $\sim$125 GeV~\cite{Collaboration:2012si,Chatrchyan:2012tx} is confirmed  by the 2012 searches at the LHC.
Then the measured Higgs production cross section times branching
ratios could certainly be reproduced by the present $t'$
model, given the parameter freedom in this model and its capability of
inducing large Higgs rate corrections of both signs.
Then investigating  this additional  Higgs production channel as the
$t'$ decay,
as discussed here, would of course be relevant in particular to
confirm the Higgs existence.
Finally, in the case of a signal from a heavy Higgs, say  
above $500$~GeV~\footnote{Such 
a heavy Higgs would be neither SM-like, as disfavored by the EW
precision tests,
nor belonging to a supersymmetric extension, as forbidden by the
Higgs sector structure.
However, it could perfectly be e.g. in a RS scenario where its
contributions to the oblique T parameter can be
compensated by new KK-induced contributions.}
as we will consider here, the same fit of Higgs data would be
instructive as a test of the present $t'$ model and similarly
the $t'$ decay should be considered as a complementary channel
of Higgs production.

\section{THE THEORETICAL MODEL}

At low scale, let us assume the presence of a unique additional
vector-like quark multiplet including a $t'$
component.
Then, irrespective of the representation of this multiplet under the
${\rm SU(2)_L}$ gauge group (i.e. the $t'$ belongs to a
singlet, doublet,\dots), the interferences between the next heavier top mass eigenstate $t_2$
[composed of $t$, $t'$, $t''$] and the $t_1$ [$\equiv$ the physical top quark whose mass is measured] 
contributions~\footnote{The top quark exchange in the loop is
the dominant contribution in the SM.}
to the triangular loop of the gluon-gluon 
fusion mechanism will be systematically constructive. This is
due to the fact that the physical signs of the Yukawa coupling
and mass insertion involved in this loop -- two chirality flips
are necessary -- will be systematically identical giving rise
to a positive product (for $t_1$ as well as for $t_2$). Hence,
the cross section of the gluon fusion mechanism 
may be increased or slightly decreased 
(because of a possible $t$ Yukawa coupling reduction) relatively to the SM case.
\\ To get the minimal scenario with only additional vector-like
quark multiplets including $t'$ components able to strongly
suppress the gluon fusion, one needs to introduce a first
top-partner $t'$ in a ${\rm SU(2)_L}$ doublet as well as
a second top-partner $t''$ in a gauge singlet. For simplification, we
do not consider the doublet including a
$b'$~\footnote{Similar results are expected in such a
situation.} that would also be exchanged
in the triangular loop. So we end up with the doublet
$(q_{5/3},t')$, $q_{5/3}$ being an exotic quark with electric
charge $5/3$
and without self-Yukawa coupling (in turn no possible loop
exchange).
Indeed, with this field content, all the possible generic mass
terms and Yukawa couplings appearing in the Lagrangian are,
\begin{equation} 
{\cal L}_{\rm Yuk.} =  
Y \overline{\left ( \begin{array}{c}
t \\  b 
\end{array} \right )}_{L}
H^\dagger t^c_R +
Y' \overline{\left ( \begin{array}{c}
q_{5/3} \\  t' 
\end{array} \right )}_L
H t^c_R +
Y'' \overline{\left ( \begin{array}{c}
q_{5/3} \\  t' 
\end{array} \right )}_{L/R}
H t''_{R/L} + 
\tilde Y \overline{\left ( \begin{array}{c}
t \\ b 
\end{array} \right )}_L
H^\dagger t''_R 
\nonumber
\end{equation}
\begin{equation} 
+ \ Y_b \overline{\left ( \begin{array}{c}
t \\  b 
\end{array} \right )}_{L}
H b^c_R 
+ m \ \bar t''_L t^c_R
+ m' \overline{\left ( \begin{array}{c}
q_{5/3} \\  t' 
\end{array} \right )}_{L}
\left ( \begin{array}{c}
q_{5/3} \\  t' 
\end{array} \right )_{R}
+ m'' \ \bar t''_{L} t''_{R}
+ {\rm H.c.}
\label{VTH:LagDoub}
\end{equation}
where $H$ represents the SM Higgs doublet and $L/R$ the fermion
chiralities.
By construction, the vector-like quarks possess same quantum
numbers and gauge group representations
for the left-handed and right-handed states.
We have not written the Yukawa couplings for the first two quark
generations as their mixings with the top-partners $t',t''$
are negligible compared to the $t$-$t'$-$t''$ mixing and the CKM
mixing angles are typically small, so that the first two quark generations
are decoupled from $b,t,t',t''$.
Note that a field redefinition rotating $t^c_R$ and $t''_R$ can
allow to eliminate the $m$ term without loss of generality.
A last remark is that the $Y''$ term could be split in two terms
with different chiralities and coupling constants.
The Lagrangian~(\ref{VTH:LagDoub}) gives rise, after EWSB, to
this top mass matrix:
\begin{equation} 
{\cal L}_{\rm mass} =
\overline{\left ( \begin{array}{c} t \\ t' \\ t'' \end{array}
\right )}_L
\left ( \begin{array}{ccc}  
Yv & 0 & \tilde Y v\\  
Y' v & m' & Y'' v\\ 
m & Y'' v & m''
\end{array} \right )
\left ( \begin{array}{c} t^c \\ t' \\ t'' \end{array} \right )_R+ {\rm H.c.}
\label{VTH:MassDoub}
\end{equation}
with $v\simeq 174$~GeV the SM vacuum expectation value of the
Higgs boson.
In our notations of Eq.(\ref{VTH:MassDoub}), the parameters $Y$,
$Y'$, $Y''$ and $\tilde Y$
contain the whole sign (i.e. the combination of the ${\rm
SU(2)_L}$ contraction signs and Yukawa coupling constant signs).
Note that vector-like fermions do not require EWSB to acquire
mass.
The non-trivial consequence of the present $t',t''$ charge
assignment choice is the presence of Yukawa terms in the
block diagonal matrix of Eq.(\ref{VTH:MassDoub}) associated to
the top-partners~\footnote{Namely the last two lines and columns
of this
mass matrix.} (such Yukawa matrix elements would be absent in
the first case of a unique top-partner). This feature of the
mass structure
allows strong suppressions of the gluon fusion mechanism. 
In particular, the own top-partner ($t',t''$) Yukawa coupling
($Y''$) sign can be chosen independently of the top ($t$) Yukawa
coupling
($Y$) sign in order to generate destructive interferences
between the top and top-partner loops.

\section{$t_2$ RATES AND DIRECT CONSTRAINTS}

We consider here the model described in the previous section, where $t'$, $t''$ denote
 the states in the interaction basis while $ t_1$, $t_2$, and $t_3$ 
stand for the mass eigenstates, with  $m_{t_3}>m_{t_2}>m_{t_1}$,    
$ t_1$ being the standard top quark and $m_{t_1}$ its physical mass.
We concentrate on the phenomenology of the next-to-lightest top mass eigenstate
$t_2$; the $t_3$ eigenstate production is sub-dominant given its larger mass. In a
second stage, one could add the contributions to the Higgs
production from the $t_3$ decays like $t_3\to t_1 h$ or $t_3\to t_2 Z$. 

In Table~\ref{VTH:DoubletTable}, we define our benchmark points
by the values of the fundamental parameters -- including the
Higgs mass $m_h$ -- and the corresponding $m_{t_2}$, $m_{t_3}$ values.
These sets of parameters are selected
in particular to have a large branching fraction $B_{t_2\to t_1 h}$ enhancing the
studied Higgs signal. Note that in the minimal model with a unique
doublet $(q_{5/3},t')$,
$B_{t_2 \to b W}$ is negligible compared to $B_{t_2 \to t_1 h}$
and $B_{t_2 \to t_1 Z}$~\cite{AguilarSaavedra:2009es}. For none
of the considered benchmark points,
the channel $t_2 \to q_{5/3} W$ is open. Table~\ref{VTH:DoubletTable} also provides
the theoretical $t_2$ widths and the $\sigma_{\bar t_2t_2}$
cross sections for the $t_2$ pair production at LHC computed
with the HATHOR program~\cite{Aliev:2010zk} at NNLO.
As a comparison, we give also in  Table~\ref{VTH:DoubletTable} the
expected SM cross
sections~\cite{LHCHiggsCrossSectionWorkingGroup:2011ti}
for Higgs production via gluon fusion, $\sigma^{\rm SM}_{\rm gg\to h}$.
It is physically important to note that the branching ratios $B_{t_2 \to t_1 h}$ and $B_{t_2 \to t_1 Z}$ are not vanishing in contrast
with the case of a fourth generation $t'$ quark so that the observation of such decays (discussed in Section \ref{VTH:exp-analysis}) 
would even prove the vector-like nature of a heavy top-like quark.

Table~\ref{VTH:DoubletTable} presents finally the CMS constraints on the
observables $\sigma_{\bar t_2t_2} B^2_{t_2}$
derived from the search for pair production of a heavy top-like quark~\cite{CERN-PH-EP-2012-081,Chatrchyan:2011ay,CMSweb}
%%% \cite{CMSweb,PhysRevLett.107.271802,CMS-PAS-EXO-11-051} 
(present bounds from ATLAS are less stringent~\cite{Aad:2012xc,ATLASweb}).
It appears that the corresponding theoretical values, predicted
in the models considered here, respect those experimental limits
for $m_{t_2}$ as low as
$\sim 400$~GeV~\footnote{We have also checked that the Tevatron
constraints are satisfied.}.
Increasing theoretically  $B_{t_2\to t_1 h}$, and consequently lowering $B_{t_2\to t_1 Z}$ and  $B_{t_2\to bW}$,  is allowed within these
constraints.

%%% Experimental constraints for Moriond 2012
%%% CMS-PAS-EXO-11-099  Title  Search for t' pair production in lepton+jets channel 2012/03/04
Due to these lower limits on $m_{t_2}$  typically around
$400$~GeV, the $t_2$ pair production suffers from a significant
phase space suppression so that
the whole rate for a single Higgs production through the $t_2$
decay is smaller than for the standard gluon fusion mechanism;
there is e.g. a factor of $\sim 10$ for 
point A1 at $14$~TeV, as shown the Table~\ref{VTH:DoubletTable}.
However, the number of Higgs events issued from the $t_2$ decay
can be significant at 14 TeV with suitable luminosities.  This Higgs production channel can thus be
an interesting Higgs boson (and $t_2$) discovery channel, among others, and especially in cases where the gluon fusion
mechanism is suppressed
by the presence of $t', t''$ states, as it occurs for instance with 
point B (see Table~\ref{VTH:DoubletTable}).

%%%%%%%%%%%%%%%%%%%%%%%%%% 
\begin{table}[!ht]
\vspace*{.5cm}
\begin{center}
\begin{tabular}{|c|c|c|c|c|c|}
\hline  
Parameter Set & A1 & A2 & B & C & D  \\
\hline 
\hline
$Y$ / $\tilde Y$   & -1.43 / 2 & 1.02 / -0.1 & 1.15 / 0.4 & 1.12 / -0.5 & 1.05 / -0.3 \\
\hline  
$Y'$ / $Y''$ & 1.85 / -1 & 1 / 0.55  & -1.5 / 1.6 & 1.1 / 1.65 & 1.7 / 1.9 \\
\hline  
$m$ / $m'$ (GeV) & 0 / 370 & 0 / 675  & 0 / 770 & 0 / 810 & 80 / 1100 \\
\hline  
$m''$ (GeV) & 510 & 645  & 980 & 850 & 1100 \\
\hline 
\hline  
$m_{t_3}$ (GeV) & 722 & 804 & 1181 & 1125 & 1454 \\
\hline  
$m_{t_2}$ (GeV) & 403 & 599  & 626 & 572 & 788 \\
\hline  
$m_{h}$ (GeV) & 125 & 125 & 255 & 320 &  540 \\
\hline  \hline  
$\sigma_{\rm gg\to h}^{\rm SM}$ (pb) @ 7 TeV & 15.31 & 15.31 & 3.18 & 2.25 & 0.58 \\
\hline 
$\sigma_{\rm gg\to h}^{\rm SM}$ (pb) @ 14 TeV & 49.85 & 49.85 & 13.50
& 10.59 & 3.85 \\
\hline
$\sigma^{\rm t'}_{\rm gg\to h}/\sigma_{\rm gg\to h}^{\rm SM}$ & 1.27 &
1.31 & 0.45 & 0.40 & 0.65 \\
\hline
$\sigma^{t'}_{\bar t_1t_1h}$ (pb) @ 7 TeV  & 0.0194& 0.0760 & 0.0037 & 0.0016 & $7 \ 10^{-4}$ \\
\hline 
$\sigma^{t'}_{\bar t_1t_1h}$ (pb) @ 14 TeV & 0.138 & 0.539 & 0.036 & 0.021 & 0.015 \\
\hline 
$\sigma_{\bar t_2t_2}$ (pb) @ 7 TeV  & 1.361& 0.0936 & 0.0709 & 0.1360 & 0.0115 \\
\hline 
$\sigma_{\bar t_2t_2}$ (pb) @ 14 TeV  & 13.53& 1.465 & 1.164 & 1.975 & 0.284 \\
\hline    
$B_{t_2 \to t_1 h}$ (\%) & 62.6 & 82.1 & 60.8 & 13.5 & 43.0 \\ 
\hline  
$B_{t_2 \to t_1 Z}$ (\%) & 28.6 & 14.7 & 25.0 & 46.1 & 40.3 \\
\hline
$B_{t_2 \to b W}$ (\%) & 8.8 & 3.2 & 14.2 & 40.4 & 16.6 \\
\hline
$\Gamma_{t_2}$ (GeV) & 4.4 & 3.5 & 19.8 & 6.5 & 8.8  \\
\hline \hline
$\sigma_{\bar t_2t_2} B^2_{t_2 \to b W}$ (pb) & 0.01 & $9 \ 10^{-5}$  & 0.001 &
0.022 & 0.000(3) \\
LHC bound~\cite{CERN-PH-EP-2012-081} & $<$ 0.26 &  $<$ 0.14 & $<$ 0.14 & $<$ 0.16  & $\times$  \\  
\hline
$\sigma_{\bar t_2t_2} B^2_{t_2 \to t_1 Z}$ (pb) & 0.11 & 0.002 & 0.004 & 0.029 & 0.002 \\
LHC bound~\cite{Chatrchyan:2011ay} & $<0.5$ & $< 0.4$ & $ <0.4$ & $< 0.4 $  & $\times$   \\  
\hline \hline
$S$ / $T$ & 0.05 / 0.05 & 0.03 / 0.03 & -0.01 / 0.23 & -0.01 / 0.30 & -0.01 / 0.28\\ 
\hline
\end{tabular}
\end{center}
\caption{Benchmark scenarios  in the present $t'$, $t''$ model, 
defined by the values of the fundamental parameters, including the Higgs mass, $m_h$, 
and the resulting $m_{t_2}$ and $m_{t_3}$ physical masses of the heavy top-like quarks. 
The  cross sections at NNLO, $\sigma_{\bar t_2t_2}$, for the $pp\to \bar t_2 t_2$ process, are shown 
at $\sqrt{s}=$ 7 TeV or 14 TeV,  together with the $t_2$ widths $\Gamma_{t_2}$ and the $t_2$ branching fraction values.
For comparison, the SM Higgs production cross section values  via gluon fusion are also given, 
together with  the ratio $\sigma^{\rm t'}_{\rm gg\to h}/\sigma_{\rm gg\to h}^{\rm SM}$, between the  
gluon fusion cross section in the present model and in the SM. 
Furthermore, the NLO cross sections for  Higgs production in association with a $\bar t t$ pair, $\sigma^{t'}_{\bar t_1t_1h}$,  
in the present model, are shown for comparison with the $\sigma_{\bar t_2t_2}$ cross sections [the $\sigma^{t'}_{\bar t_1t_1h}$ values are too small to
contribute to the signal, analyzed in this paper, from  ${\bar t_2t_2}$ production, with $t_2\to th$ decay].
Finally, the LHC upper limits and theoretical predictions
for the observables $\sigma_{\bar t_2t_2} B^2_{t_2 \to b W}$,
$\sigma_{\bar t_2t_2} B^2_{t_2 \to t_1 Z}$ are shown just above
the last line
(the crosses indicate the absence of experimental limit at the
associated $m_{t_2}$ values). In the last line the values of the 
oblique parameters $S$ and $T$ are given [after subtraction of the SM
contributions to include only new physics effects].
For the point A1 (A2), we have used the indicated $Y''$ value for the
$t''_R$ coupling and $Y''=-0.3$ ($-1.75$) for the $t''_L$ vertex.}
\label{VTH:DoubletTable}
\end{table}

\section{CONSTRAINTS FROM THE DIRECT HIGGS BOSON SEARCH}

\noindent Sets {\bf A1/A2} - At $m_{h}=125$~GeV, all the sensitive
channels for searching the Higgs boson at hadron colliders are
the decays $h\to \gamma\gamma$, $h\to WW$ (with $W\to \ell\nu$),
$h\to ZZ$ (with $Z\to \ell\bar \ell$), $h\to\tau\bar \tau$ and
$h\to b\bar b$.
The latest bounds on the Higgs boson rates obtained at the LHC
read as,
$\sigma_{\rm gg\to h}B_{\rm h\to \gamma\gamma}/\sigma_{\rm gg\to
h}^{\rm SM}B_{\rm h\to \gamma\gamma}^{\rm SM}$
$\lesssim 2$, 
$\sigma_{\rm gg\to h}B_{\rm h\to WW}/\sigma_{\rm gg\to h}^{\rm
SM}B_{\rm h\to WW}^{\rm SM}$
$\lesssim 1.30$, 
$\sigma_{\rm gg\to h}B_{\rm h\to ZZ}/\sigma_{\rm gg\to h}^{\rm
SM}B_{\rm h\to ZZ}^{\rm SM}$
$\lesssim 2.2$,  
$\sigma_{\rm gg\to h}B_{\rm h\to \tau\tau}/$
$\sigma_{\rm gg\to h}^{\rm SM}B_{\rm h\to \tau\tau}^{\rm SM}$
$\lesssim 3.2$ and
$\sigma_{\rm gg\to h}B_{\rm h\to bb}/\sigma_{\rm gg\to h}^{\rm
SM}B_{\rm h\to bb}^{\rm SM}$
$\lesssim 3.2$ \cite{CMSweb,ATLASweb,CERNcouncil,Collaboration:2012si,ATLASnote019,Chatrchyan:2012tx}. 
These bounds are compatible with the rates calculated taking
into account the $t-t'-t''$ mixing effect on the top quark
Yukawa coupling as
well as the $t_2$ and $t_3$ eigenstate contributions in the
triangular loop of the gluon fusion mechanism,
for our parameter sets A1/A2. This parameter set yields indeed:  
\begin{equation}
\sigma_{\rm gg\to h}^{\rm t'}B_{\rm h\to \gamma\gamma}^{\rm t'}/\sigma_{\rm gg\to h}^{\rm SM}B_{\rm h\to \gamma\gamma}^{\rm SM}=1.16 \ (1.19) \ \ ; \ \
\sigma_{\rm gg\to h}^{\rm t'}B_{\rm h\to WW}^{\rm t'}/\sigma_{\rm gg\to h}^{\rm SM}B_{\rm h\to WW}^{\rm SM}=1.25 \ (1.28)
\label{VTH:SigB1}
\end{equation}
\begin{equation}
\sigma_{\rm gg\to h}^{\rm t'}B_{\rm h\to ZZ}^{\rm t'}/\sigma_{\rm gg\to h}^{\rm SM}B_{\rm h\to ZZ}^{\rm SM}=1.25 \ (1.28) \
\ ; \ \
\sigma_{\rm gg\to h}^{\rm t'}B_{\rm h\to \tau\tau}^{\rm t'}/\sigma_{\rm gg\to h}^{\rm SM}B_{\rm h\to \tau\tau}^{\rm SM}=1.25 \ (1.28)
\label{VTH:SigB2}
\end{equation}
\begin{equation}
\sigma_{\rm gg\to h}^{\rm t'}B_{\rm h\to bb}^{\rm t'}/\sigma_{\rm gg\to h}^{\rm SM}B_{\rm h\to bb}^{\rm SM}=1.25 \ (1.28) \ \ \
\ [A1] \ ([A2] )
\label{VTH:SigB3}
\end{equation}
The cross section for the Higgs production is enhanced,
$$\sigma^{\rm t'}_{\rm gg\to h}/\sigma_{\rm gg\to h}^{\rm SM}=
1.27 \ [A1] \ (1.31 \ [A2] ),$$
due to the combination of two possible effects: 
the increase of the $t_1$ Yukawa coupling and the constructive
interferences between the $t_1$ contribution and the $t_2$,
$t_3$ ones.
In contrast, the branching fraction  for the decay channel into
diphoton is slightly decreased,
$$B_{\rm h\to \gamma\gamma}^{\rm t'}/B_{\rm h\to
\gamma\gamma}^{\rm SM}= 0.91 \ [A1] \ (0.90 \ [A2] ).$$ But the resulting
product
$\sigma_{\rm gg\to h}^{\rm t'}B_{\rm h\to \gamma\gamma}^{\rm
t'}$ is increased relatively to the SM case as shown in 
Eq.(\ref{VTH:SigB1}).
\\ Such an increased value of the observable
$\sigma_{\rm gg\to h}B_{\rm h\to \gamma\gamma}$, induced here by the presence of $t'$ quarks ({\it c.f.} Eq.(\ref{VTH:SigB1})),
could be indicated by the slight excess in the ATLAS \cite{ATLASweb,Collaboration:2012si,ATLASnote019,CERNcouncil} 
and CMS~\cite{CMSweb,Chatrchyan:2012tx,CERNcouncil} 
data correspon\-ding to a possible $\sim$125 GeV Higgs signal in the diphoton channel. All the values of the quantities,
$\sigma_{\rm gg\to h}^{\rm t'}B_{\rm h\to \gamma\gamma}^{\rm t'}/\sigma_{\rm gg\to h}^{\rm SM}B_{\rm h\to \gamma\gamma}^{\rm SM}$, 
$\sigma_{\rm gg\to h}^{\rm t'}B_{\rm h\to WW}^{\rm t'}/\sigma_{\rm gg\to h}^{\rm SM}B_{\rm h\to WW}^{\rm SM}$, 
$\sigma_{\rm gg\to h}^{\rm t'}B_{\rm h\to ZZ}^{\rm t'}/\sigma_{\rm gg\to h}^{\rm SM}B_{\rm h\to ZZ}^{\rm SM}$,
$\sigma_{\rm gg\to h}^{\rm t'}B_{\rm h\to \tau\tau}^{\rm t'}/\sigma_{\rm gg\to h}^{\rm SM}B_{\rm h\to \tau\tau}^{\rm SM}$ and 
$\sigma_{\rm gg\to h}^{\rm t'}B_{\rm h\to bb}^{\rm t'}/\sigma_{\rm gg\to h}^{\rm SM}B_{\rm h\to bb}^{\rm SM}$, in Eq.(\ref{VTH:SigB1})-(\ref{VTH:SigB3}),
are also compatible with the CMS and ATLAS best-fit values whose central value, from the combination of all search channels, is of  
$1.22^{+0.31}_{-0.39}$~\cite{Chatrchyan:2012tx} (for a Higgs mass hypothesis of $124$~GeV)~\footnote{This number has just been updated to $0.94^{+0.26}_{-0.34}$ for a $125$~GeV Higgs~\cite{MoriondCMSHiggs}.}  
and $0.90^{+0.40}_{-0.37}$~\cite{ATLASnote019} (for a Higgs mass hypothesis of $126$~GeV), respectively~\footnote{The present parameters lead to a Higgs rate, for the $WW$ channel, standing slightly outside the $1\sigma$ 
range but the experimental error bars obtained by both the ATLAS and CMS Collaborations are significant.}.     
\\ It is also interesting to note that the present
theoretical model allows for either an increase of $\sigma^{\rm
t'}_{\rm gg\to h}$
compared to the SM, as here for A1/A2, or a decrease as with the parameter sets in the following.

\noindent Sets {\bf B,C} - For these sets of parameters where
$m_{h}=255$~GeV or $320$~GeV, all the Higgs decays have
negligible
widths relatively to the dominant channels $h\to ZZ$ and $h\to
WW$, as in the SM case.
Hence the branching fractions $B_{\rm h\to ZZ}$ and $B_{\rm h\to
WW}$ remain unchanged in the present model
with vector-like top quarks where only the decay widths for
$h\to t\bar t$, $h\to gg$, $h\to \gamma \gamma$ and $h\to \gamma
Z$ are modified.
In consequence, the experimental limits on $\sigma_{\rm gg\to 
h}/\sigma_{\rm gg\to h}^{\rm SM} \lesssim 0.45$ ($0.40$) [for $m_{h}\simeq 255
\ (320)$~GeV]
issued from the LHC combined investigations using the $h\to
ZZ,WW$ channels exclusively
\cite{CMSweb,ATLASweb,Chatrchyan:2012tx,Collaboration:2012si,ATLASnote019,CERNcouncil} can be applied directly to our framework
where one gets
\begin{equation}
\sigma_{\rm gg\to h}^{\rm t'}/\sigma_{\rm gg\to h}^{\rm SM}
=0.45 \ \ [B] \ \ ; \ \ 0.40 \ \ [C]
\end{equation}
which does not conflict with the above LHC limits.
\\ Note that for the point C, $\sigma_{\rm gg\to h}^{\rm t'}$ is
strongly reduced compared to SM. A factor $1/10$ could even be
achieved in the
present theoretical model but variants of the multiplet choice
(non-minimal in term of field content),
allowing coupling correction cancellations, should 
then be used instead to pass the indirect constraints discussed
in Section~\ref{VTH:oblique}. 

\noindent Set {\bf D} - For  $m_{h}=540$~GeV, the Higgs boson is
searched only through its decays into $ZZ$ and $WW$.
The strongest bounds on the Higgs rates from the LHC read as
$\sigma_{\rm gg\to h}B_{\rm h\to ZZ}/\sigma_{\rm gg\to h}^{\rm SM}B_{\rm h\to ZZ}^{\rm SM} \lesssim 0.90$ and
$\sigma_{\rm gg\to h}B_{\rm h\to WW}/\sigma_{\rm gg\to h}^{\rm
SM}B_{\rm h\to WW}^{\rm SM} \lesssim 1.45$~\cite{CMSweb,ATLASweb,CERNcouncil,Collaboration:2012si,ATLASnote019,Chatrchyan:2012tx}.
These upper limits are clearly in good agreement with the rates
calculated in the presence of the $t'$ and $t''$ states (that
modifies
$B_{\rm h\to t\bar t}$) for the set D, namely,
\begin{equation}
\sigma_{\rm gg\to h}^{\rm t'}B_{\rm h\to ZZ}^{\rm
t'}/\sigma_{\rm gg\to h}^{\rm SM}B_{\rm h\to ZZ}^{\rm SM} = 0.69
\ \ \ \ \ \
\sigma_{\rm gg\to h}^{\rm t'}B_{\rm h\to WW}^{\rm
t'}/\sigma_{\rm gg\to h}^{\rm SM}B_{\rm h\to WW}^{\rm SM} = 0.69
\\
\end{equation}
where $$B_{\rm h\to ZZ}^{\rm t'}/B_{\rm h\to ZZ}^{\rm SM} = 1.06
\ \ \ \ \ \ B_{\rm h\to WW}^{\rm t'}/B_{\rm h\to WW}^{\rm SM} =
1.06 \ \ [D].$$

\section{THE INDIRECT CONSTRAINTS AND OBLIQUE PARAMETERS}
\label{VTH:oblique}

Given the absence of precise measurement for the $Zt\bar t$
vertex (coupling directly modified by the $t-t'-t''$ mixing),
the main indirect constraints to the present model come from the corrections to the gauge boson vacuum
polarizations induced by the loops of $q_{5/3}$,$t'$,$t''$
states. The values of the oblique parameters $S,T$
that we obtain, according to the preliminary calculations of
Ref.~\cite{Barbieri:2006bg,Lavoura:1992np}, are given in
Table~\ref{VTH:DoubletTable}.
They appear to belong to the $1\sigma$ regions induced by the
long list of EW precision observables measured mainly at the LEP
collider
\cite{Nakamura:2010zzi}. 
\\ Remark that the input parameters of
Table~\ref{VTH:DoubletTable} (i.e. the theoretical values in the
first four lines) have been chosen to
fix a panel of characteristic benchmark points for $m_{t_2}$
that pass the indirect constraints as well as the bounds from
direct Higgs search described in previous section; however those
two types of constraints allow large domains of the parameter
space (varying also $m_h$). The precise setting of the $Y$
coupling reflects mainly
the experimental precision on the top quark mass measurement (and not any fine-tuning).

The $t',t''$ states could contribute to Flavor Changing Neutral Current (FCNC) reactions which are experimentally well constrained; from the theoretical point of view,  
these FCNC contributions rely precisely on the whole set of Yukawa coupling constants for the entire quark sector. The treatment of such an high degree of freedom 
in the parameter space is beyond the scope of the present study.

Finally, given the relative heavyness of the $t_2$ quark, we have checked that the predicted value for the $V_{tb}$ CKM matrix element 
is in agreement with the experimental measurement close to unity obtained (without assuming $3\times 3$ unitarity) through the single top
quark production cross section at Tevatron \cite{Nakamura:2010zzi}.

%%%-------------------------------------------------------------------------------------------
%%%---------------- MF Revised 20120305 BEGIN -----------------------------------------------
%%%-------------------------------------------------------------------------------------------
\section{HIGGS SIGNAL RECONSTRUCTION IN $\bar{t}_2 t_2\rightarrow th+X$ EVENTS}
\label{VTH:exp-analysis}

We have studied the sensitivity at the LHC with $\sqrt{s}=14$~TeV of a search for $pp \to t_2 \bar t_2$ production, with one of the $t_2$
decaying to $th$, and the other decaying to $bW$ or $tZ$ or $th$, resulting into $thbW$, $thtZ$ and $thth$ final states, respectively. 
For a best sensitivity and in order to measure the Higgs mass precisely, we have exploited
the Higgs decay channel into $ZZ$ to four charged leptons for the signals with Higgs mass above $200$~GeV (points B, C, D) 
while the Higgs decay into two photons is considered for the signals with a Higgs mass of $125$~GeV (sets A1 and A2). 

For the signal event generation, we have implemented the couplings of our $t_2$ model in {\small \tt FeynRules}~\cite{Christensen:2008py}~\footnote{We thank Claude
Duhr and Benjamin Fuks for their precious help in this
implementation.}
interfaced with {\small \tt MadGRAPH}~\cite{Herquet:2008zz} for
the {\it Monte Carlo} generation, {\small \tt
PYTHIA}~\cite{Sjostrand:2006za}
for the hadronization part and {\small \tt
DELPHES}~\cite{Ovyn:2009tx} for the fast simulation of a typical LHC detector response.
Signal events are generated for the $t_2$ and $h$ masses corresponding to the parameter sets described in Table~\ref{VTH:DoubletTable}, for the three final states $thbW$, $thtZ$ and  $thth$. Events corresponding to points A1 and A2 are generated with the Higgs decaying through $h\to\gamma\gamma$, while for points B, C and D the $h\to ZZ$ decay is retained for the sensitivity studies described in this Section. 
The main backgrounds were generated with {\small \tt ALPGEN}~\cite{Mangano:2002ea} interfaced to {\small \tt PYTHIA}
and {\small \tt DELPHES}, as for the signal events. Physics
objects used for the analysis (photons, leptons and jets) were 
defined emulating the requirements used in real CMS Higgs
searches in the 2011 data. In particular, we followed closely the physics object definition as for the  real data 7 TeV Higgs analysis in the diphoton channel~\cite{Chatrchyan:2012tw} %%% {CMS-PAS-HIG-11-030}
and in the four lepton channel~\cite{Chatrchyan:2012dg}. %%%{CMS-PAS-HIG-11-025}. 
% Reconstructed jets  are required to have a transverse momentum, $p_T$, larger than 30 GeV, 
% while for leptons (electrons or muons) and  photons a transverse momentum lower cut of 20 GeV is applied.

In the following the quoted number of events and distributions are normalized to an integrated luminosity of $20$ fb$^{-1}$ multiplied by, for the different signal final states,  their expected $\sigma_{\bar t_2t_2}$ cross section times branching fractions (hereafter referred to as signal event yield per unit of integrated luminosity), while, for background processes,  their {\small \tt ALPGEN} cross sections are used. 
The signal event yield in the different final states depends upon the parameter set under consideration. 
In Table~\ref{VTH:signalyield}, we summarize, for each parameter set described in Table~\ref{VTH:DoubletTable}, the physical signal $m_{t_2}$ and $m_h$ masses, the $\sigma_{\bar t_2t_2}$ cross sections, the $t_2$ branching fractions into $bW$ ($B_{t_2\rightarrow bW}$), $tZ$ ($B_{t_2\rightarrow tZ}$), $th$ ($B_{t_2\rightarrow th}$)  final states, as well as the $h$ branching fractions into $\gamma\gamma$ (for sets A1 and A2) and into $ZZ$ (for sets B, C and D), reporting the expected SM branching fraction values and the factor $f_{h\to VV}^{t'/SM}$ (see Section IV) by which it is modified in the present model. The last three columns of Table~\ref{VTH:signalyield} show the expected signal event yield,  for  $thbW$, $thtZ$ and $thth$ final states, with one Higgs  boson decaying via  to diphoton or ZZ  final state.  
They are calculated as follows: 
\begin{center}
\begin{tabular}{ll}
$thbW$ final state, &\hspace{0.2cm} $Y_S(h\to VV) =\ 2\ B_{t_2\rightarrow th} B_{t_2\rightarrow bW}\  B^{SM}_{h\to VV} f_{h\to VV}^{t'/SM} \  \sigma_{\bar{t}_2t_2}$ \\
$thtZ$  final state, & \hspace{0.2cm} $Y_S(h\to VV) =\ 2\ B_{t_2\rightarrow th} B_{t_2\rightarrow tZ}\  B^{SM}_{h\to VV} f_{h\to VV}^{t'/SM} \  \sigma_{\bar{t}_2t_2}$\\
$thth$  final state,& \hspace{0.2cm} $Y_S(h\to VV) =\ 2 \ B^2_{t_2\rightarrow th} \ B^{SM}_{h\to VV} f_{h\to VV}^{t'/SM}\  \sigma_{\bar{t}_2t_2}$\\
\end{tabular}
\end{center}

For $thth$ events, the event yield is given for events where one $h$ decays to vector bosons, while the second $h$ decays inclusively. The signal event yield (in fb) multiplied by the integrated luminosity (in fb$^{-1}$) results into the numbers of produced signal events for that luminosity.  
We note that, for a chosen Higgs decay,  the signal yield depends on  both $\sigma_{\bar t_2t_2}$ and $B_{t_2\rightarrow th}$. 
Then, for example, in spite of the fact that  $\sigma_{\bar t_2t_2}$  is larger for point C than B, the signal event yield is higher for point B than C, because of the larger $B_{t_2\rightarrow th}$ value in  point B.  The background cross sections are listed  in Tablel~\ref{VTH:tabzzyields}, for the four lepton search channel, and in Table~\ref{VTH:diphoyield} for the diphoton search channel.
%%% C:\Documents and Settings\marta\My Documents\Fromhp07MartaNicePrivate\MartaNicePrivate\2012\2012-VT2H-publication\notes\20120320-xsecbrnumders 
\begin{table}[!ht]
\begin{center}
\begin{tabular}{|c|c|c|c|c|c|c|c|c|r|r|r|}
\hline   
Parameter set &  $m_{t_2}$(GeV) & $m_h$(GeV)  &  $\sigma_{\bar t_2t_2}$(fb) & $B_{t_2\rightarrow th}$  &  $B_{t_2\rightarrow tZ}$  & $B_{t_2\rightarrow bW}$ &  $B^{SM}_{h\to VV}$ & $f_{h\to VV}^{t'/SM}$ & \multicolumn{3}{c|}{ $Y_S(h\to VV)$(fb)\ } \\  \cline{10-12}
 & & & & & & & & & \hspace{0.5cm}$thbW$ & \hspace{0.5cm}$thtZ$ & \hspace{0.5cm}$thth$  \\  \hline
Point A1 &403 & 125 & $1.353\ 10^{4}$ & 0.626 & 0.286 & 0.088 & $2.29\ 10^{-3}$ & 0.91 & 3.11 & 10.1 & 22.1 \\
Point A2 & 599 &125& $1.465\ 10^{3}$& 0.821& 0.147& 0.032&  $2.29\ 10^{-3}$&  0.90&  0.159  & 0.729  & 4.07 \\
Point B   & 626 & 255 & $1.164 \  10^{3}$& 0.608 & 0.250 & 0.142 & 0.298 & 1.00 & 59.9   &105  & 256 \\
Point C  & 572 &320 & $1.975\ 10^{3}$ & 0.135 & 0.461 & 0.404 & 0.309 & 1.00 & 66.6  & 76.0  & 22.2 \\
Point D  & 788 & 540 & $0.284\ 10^{3}$ & 0.430 & 0.403 & 0.166 & 0.265 & 1.06 & 11.4  &  27.6  &  29.5\\
\hline
\end{tabular}
\caption{Information entering the signal event yield $Y_S$ calculation. 
The $t_2$ pair production cross section at 14 TeV and  $t_2$ branching fraction values are from Table ~\ref{VTH:DoubletTable}. The  ratio $f_{h\to VV}^{t'/SM}=B^{t'}_{h\to VV}/B^{SM}_{h\to VV}$ values are calculated in Section IV.
The event yield $Y_S$, per unit of integrated luminosity, is given  for the different final states $\bar t_2  t_2\to thbW,\ thtZ,\ thth$ and $h\to\gamma\gamma$, for points A1 and A2, or  $h\to ZZ$, for points B, C, D. For $thth$ events, the event yield is calculated for the case where one $h$ decays to vector bosons, while the second $h$ decays inclusively. This synoptic table summarizes at once, for convenience, all the relevant numbers useful
for the experimental search.}
\label{VTH:signalyield}
\end{center}
\end{table}

%%%-----------------------------------------------------------------------------------------------------------------------
\subsection{Search for $\bar{t}_2 t_2\rightarrow th+X$ signal in the  four leptons plus multijets channel}
%%%----------------------------------------------------------------------------------------------------------------------- 
In order to estimate the sensitivity of a search for $thbW,\ thtZ$ and $thth$ final states, when the Higgs boson is relatively heavy, as expected for the points B, C and D, with $m_{h}=255, 320, 540$~GeV, respectively, we exploit 
the decay channel into four charged leptons $h\to ZZ\rightarrow 4l$. Signals from $thbW,\ thtZ,\ thth$, $h\to ZZ\to 4l$, final states are characterized by four high transverse momentum leptons, from the Higgs decay, and a large number of energetic jets, from accompanying top and heavy vector boson decays.

The event selection, exploiting the large number of high transverse momentum ($p_T$) leptons and jets, as well as the b-jet content of the event, consists of the 
following criteria:
\begin{itemize}
\item Four leptons (muons or electrons) are required with transverse momentum $p_T>20$~GeV and pseudo-rapidity $|\eta|<$~2.4 or 2.5 for muons or electrons, respectively, and  two lepton pairs, each pair with same flavor but opposite charge leptons. The lepton pair of highest 
$p_T$ is required to have a dilepton invariant mass consistent with the Z mass, $M_{2l}=M_Z\pm15$ GeV, while the lepton pair with second highest dilepton $p_T$ must have a dilepton invariant mass $M_{2l}>12$ GeV;
\item 
$H_T>$ 1000 GeV and $N_j>$ 5, with $H_T$ being defined as the scalar sum of the transverse momenta of identified  leptons, photons, jets, and the missing energy, and  $N_j$ being the number of hadronic jets with $p_T>$30 GeV and $|\eta|<$2.4;
%%% should use 50 GeV jets to make sure the btag efficiency values apply - it does not cut on signal and not on background numbers after HT>1000 GeV - but will change the Nj distribution, for ttbar
\item  At least two b-tagged jets in the event. 
\end{itemize}
%%%-------------------------------
\begin{figure}[htbp]
\begin{center}
\begin{tabular}{cc}
\includegraphics[width=0.5\textwidth]{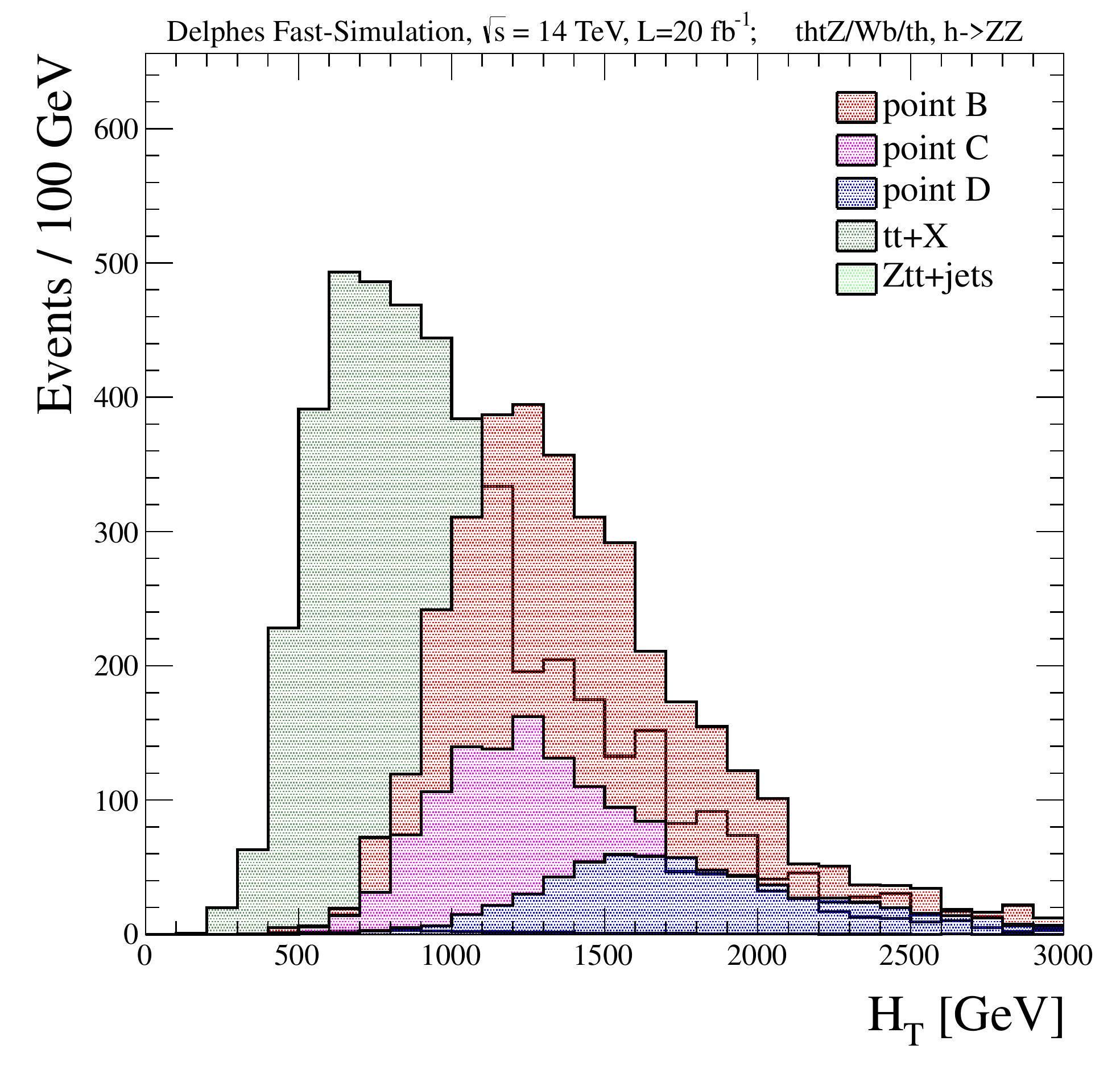}
&
\includegraphics[width=0.5\textwidth]{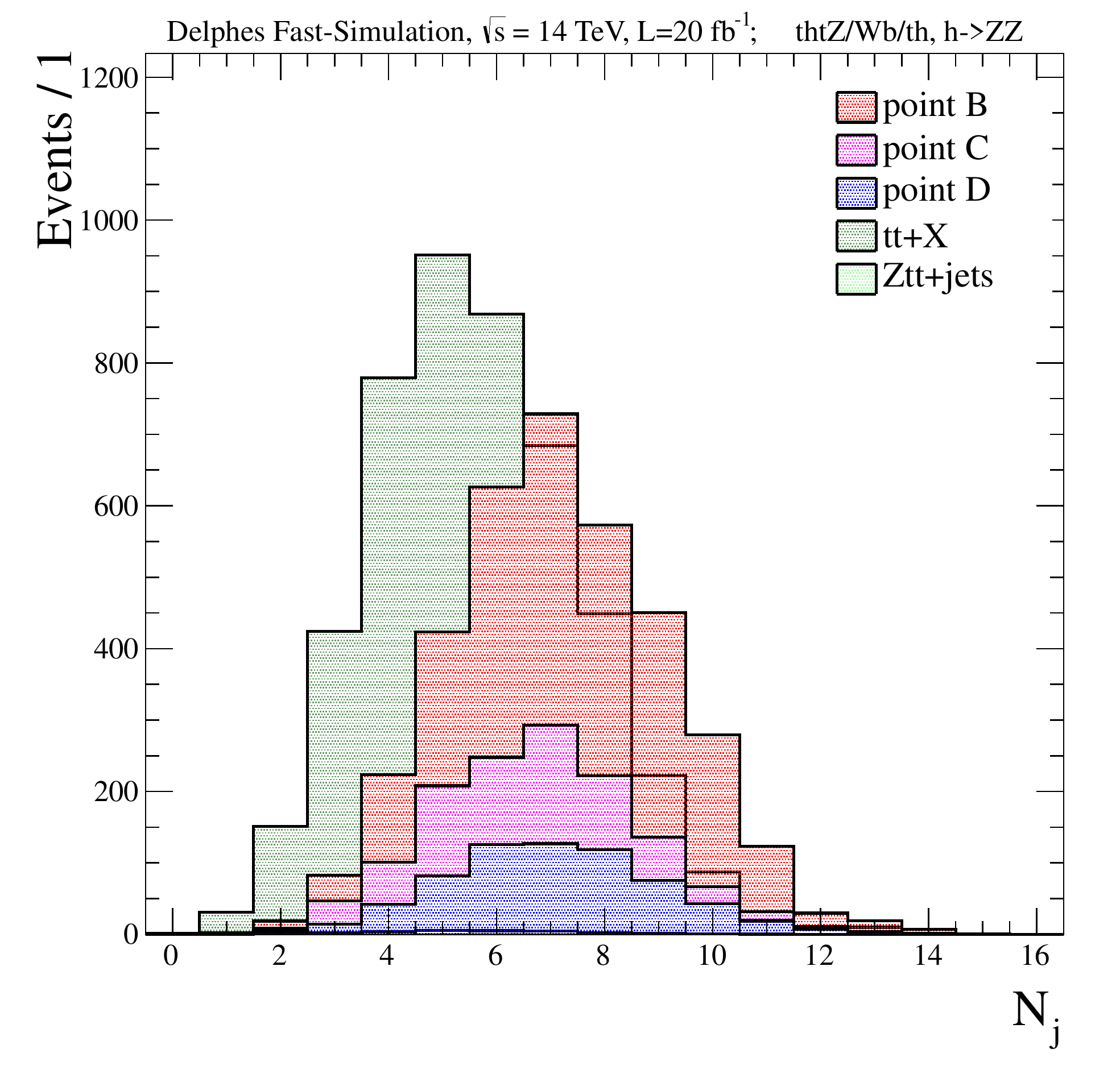}
\\
\end{tabular}
\caption{Distributions of the $H_T$ variable and the jet multiplicity $N_j$ after four lepton requirements. Background and signal distributions are shown overlaid, to compare the different distribution shapes, with the background normalized to the expected events for 20~fb$^{-1}$ and the corresponding signal expectation multiplied by a factor of fifty. }
\label{VTH:ZZnjets}
\end{center}
\end{figure}

The motivation for the $H_T$ and $N_j$ cuts is evident from  Figure~\ref{VTH:ZZnjets}, showing the  $H_T$ and $N_j$ distributions after the four lepton requirements. The event numbers are shown for  an integrated luminosity $L=20$ fb$^{-1}$, with the signal event numbers multiplied by a factor of fifty, for shape comparison.
Expectations for the signals, by final state,  and the backgrounds processes, after the selection cuts, for an integrated luminosity of  $L=20$ fb$^{-1}$, are shown in Table~\ref{VTH:tabzzyields}, where the following quantities are reported:
\begin{itemize}
\item $Y_S$, the signal event number produced,  per unit of integrated luminosity, given for each signal parameter set and final state under study,  as reported in Table~\ref{VTH:signalyield};
\item $\sigma$, the cross sections, as calculated by the ALPGEN generator,  for the background processes included in the present study; 
\item $N_{4l}$, the number of events after four lepton requirements;
\item $N_{N_j,H_T}$, the number of events after the additional requirements $H_T>$ 1000 GeV and $N_j>$ 5;
\item $N_{2b}$, the number of events after the additional requirements of two b-tagged jets in the event;
\item $S$ and $B$, the total number of signal (summed over all signal final states) and background (summed over all background processes),
after a cut on the reconstructed four lepton invariant mass, $M_{4l}$,  with $\Delta B$ being the expected uncertainty on the total number of background events. Here we only consider the statistical uncertainty  $\Delta B=\sqrt{B}$, but a more realistic estimation should also include systematic uncertainties on the number of background events. Some of the dominant systematic uncertainties are depending on detector effects (for instance jet energy resolution) and can be best evaluated with real data.   In the following, an estimation of the signal sensitivity is given as the ratio  $S/\Delta B$, measuring the signal in terms of background standard deviations.
\end{itemize}
%%%------------------------------------------------
\begin{table}[htbp]
\begin{center}
\begin{tabular}{|c|c|c|c|c|c|r|r|r|}
\hline
        & \multicolumn{4}{c|}{Signal: $thbW / thtZ / thth$, $h\to ZZ$ } &  \multicolumn{4}{c|}{Total  signal $S$, $M_{4l}$(GeV) cut}\\ \hline
 Parameter set & $Y_S$ (fb) &  $N_{4l}$ & $N_{N_j,H_T}$ & $N_{2b}$ & no $M_{4l}$ cut & \hspace{0.2cm}$M_{4l}>200$& \hspace{0.2cm}$>300$ &  \hspace{0.2cm}$>500$ \\ \hline
 Point B                   & 59.9 /105 /256    & 8.1 /23.7 /37.8        &    5.5 /18.0 /28.9   &  3.52 /11.5 /18.5& 33.4 & 30.4 &   17.0  & 4.9\ \\  
\hline  
 Point C                   & 66.6 /76.0 /22.2  &  7.7 /15.8 /3.1        &    4.8 /11.3 /2.2      &  3.07 /7.23 /1.41&    11.7 & 10.7    & 7.4  &2.0\\       
\hline
%%% point D corrected with fact 1.06
 Point D                   & 11.4 /27.6 /29.5   &  1.5 /6.6 /4.7         &  1.1 /5.3 /3.9          &  0.70 /3.40 /2.50&      6.6 & 6.2    &    5.1  &2.7\\    
\hline
         &  \multicolumn{4}{c|} {Background }& \multicolumn{4}{c|}{Total background $B$,  $M_{4l}$(GeV) cut}\\ \hline
 Process  & $\sigma$(fb)          &   $N_{4l}$ & $N_{N_j,H_T}$ & $N_{2b}$                     & no $M_{4l}$ cut  & $M_{4l}>200$ & $>300$ & $>500$\\ \hline
%%%--------------------------------------------------------------------------------- 2b rej.fac.=0.074 x 1480 = 109
%%%---------------------------------------------------------------------------------------------- multiply by rej fac NM4l>200/N5J=0
%%%                                                                                                                               from Tab V and VI 20120323
$t\bar{t}$ + jets                 &   9.19$\  10^{5}$  &    4680     &  1480          &     27.5      & 35.5      &   23.5     & 13.2       & 6.8\\
\cline{6-9}
%%%------------------------------------------------------------------ treat like signal fac = 0.81
$t\bar{t}b\bar{b}$ + jets   &   2.50$\  10^{3}$  &     5.60        &   3.10      &           2.0    &      \multicolumn{4}{c|}{Statistical $\Delta B$, $M_{4l}$(GeV) cut}\\ \cline{6-9}
%%%------------------------------------------------------------------ treat like 1b = 0.1
$t\bar{t}W$ + jets              &   1.99$\  10^{2} $ &      1.20      &    0.40      &          0.036    & no $M_{4l}$ cut &$M_{4l}>200$ &$>300$ &$>500$\\ \cline{6-9}
%%%------------------------------------------------------------------ treat like signal fac = 0.81
$t\bar{t}Z$ + jets               &   97.3$\ $          &    25.0      &    9.5          &          6.0        &  6.0 & 4.9  & 3.6  & 2.6 \\
\hline
\end{tabular}
\end{center} 
\caption{Cross section and number of expected events after different selection cuts (see text), for an integrated luminosity of 20~fb$^{-1}$ at $\sqrt{s}=14$ TeV,  in the search for $thbW,$ $thtZ$ and  $thth$ final states, with $h\rightarrow ZZ$, in the four leptons plus multijets channel.  
The number of events is given separately for each signal final state ($thbW / thtZ / thth$) and for each background process. 
In the four rightmost columns, 
$S$, the total number of signal events (summed over all signal final states),  and $B$, the total number of background events 
(summed over all background processes), are given after all selection cuts but the $M_{4l}$ cut, in column ``no $M_{4l}$ cut",  and after three different $M_{4l}$ cut values. 
The $S$ values for points B, C and D are given on the same lines as indicated in column ``Parameter set". 
The $B$ values in the column ``no $M_{4l}$ cut"  are the sum of the individual background contributions as reported in the background column under ``$N_{2b}$".   
The quantity $\Delta B=\sqrt{B}$ is  the expected statistical uncertainty on the total number of background events. In the text, the signal sensitivity is given as the ratio  $S/\Delta B$.}
\label{VTH:tabzzyields}
\end{table}

The need for additional cuts, after the leptons, jets and $H_T$ requirements, is seen in Table~\ref{VTH:tabzzyields}, where the $N_{N_j,H_T}$ column shows that after these cuts, a large background, predominantly from $t\bar{t}$ + jets production, is still present. 
Further background rejection can be obtained with requirements on the number of b-tagged jets. 
The motivation is the following: in  $t\bar{t}$ events the requirement of  four high $p_T$ leptons strongly reduces the number of b-taggable jets. Indeed, in $t\bar{t}$+jets events,   while two  high $p_T$ leptons are provided by the two leptonic W decay, the additional two leptons are preferentially resulting from the two $b$ decays.  Thus, applying a $b$-tagging requirement can reduce  significantly the $t\bar{t}$ 
contribution. In the signal case,  the four leptons come mainly from the $h\to ZZ\to 4l$ decay so that there are in average at least two $b$-taggable jets in the event, for which a $b$-tagging requirement can have high efficiency. 
Since no detailed b-tagging information is presently available in {\small \tt DELPHES}, we make use of known $b$-tagging and mis-tagging efficiency values as measured by LHC experiments (see {\it e.g.} Ref.~\cite{CMS-PAS-BTV-11-002}),
%%% see eg CMS btag measurements   b-Jet Identification in the CMS Experiment -CMS-PAS-BTV-11-004, 2012/02/27, 
%%% http://cdsweb.cern.ch/record/1427247/files/BTV-11-004-pas.pdf
%%% see eg CMS btag measurements   Status of b-tagging tools for 2011 data analysis - CMS-PAS-BTV-11-002, 2011/11/03
%%% http://cdsweb.cern.ch/record/1395489/files/BTV-11-002-pas.pdf
to estimate the effect on background and signal of requiring two b-tagged jets in the event. 
The probability for a $b$-jet to be $b$-tagged is taken to be $\epsilon_{b-tag}=0.8$ and for a mis-tagging the probability is  $\epsilon_{mis-tag}=0.05$. An event-by-event weight, that is an estimation of the probability that two $b$-jets be identified in the event, is applied to both background and signal events. The number of expected events, for signals and backgrounds, resulting from  leptons, jets and b-tagging requirement are shown in the column $N_{2b}$ of Table~\ref{VTH:tabzzyields}.

\begin{figure}[htbp]
\begin{center}
\begin{tabular}{cc}
\includegraphics[width=0.5\textwidth]{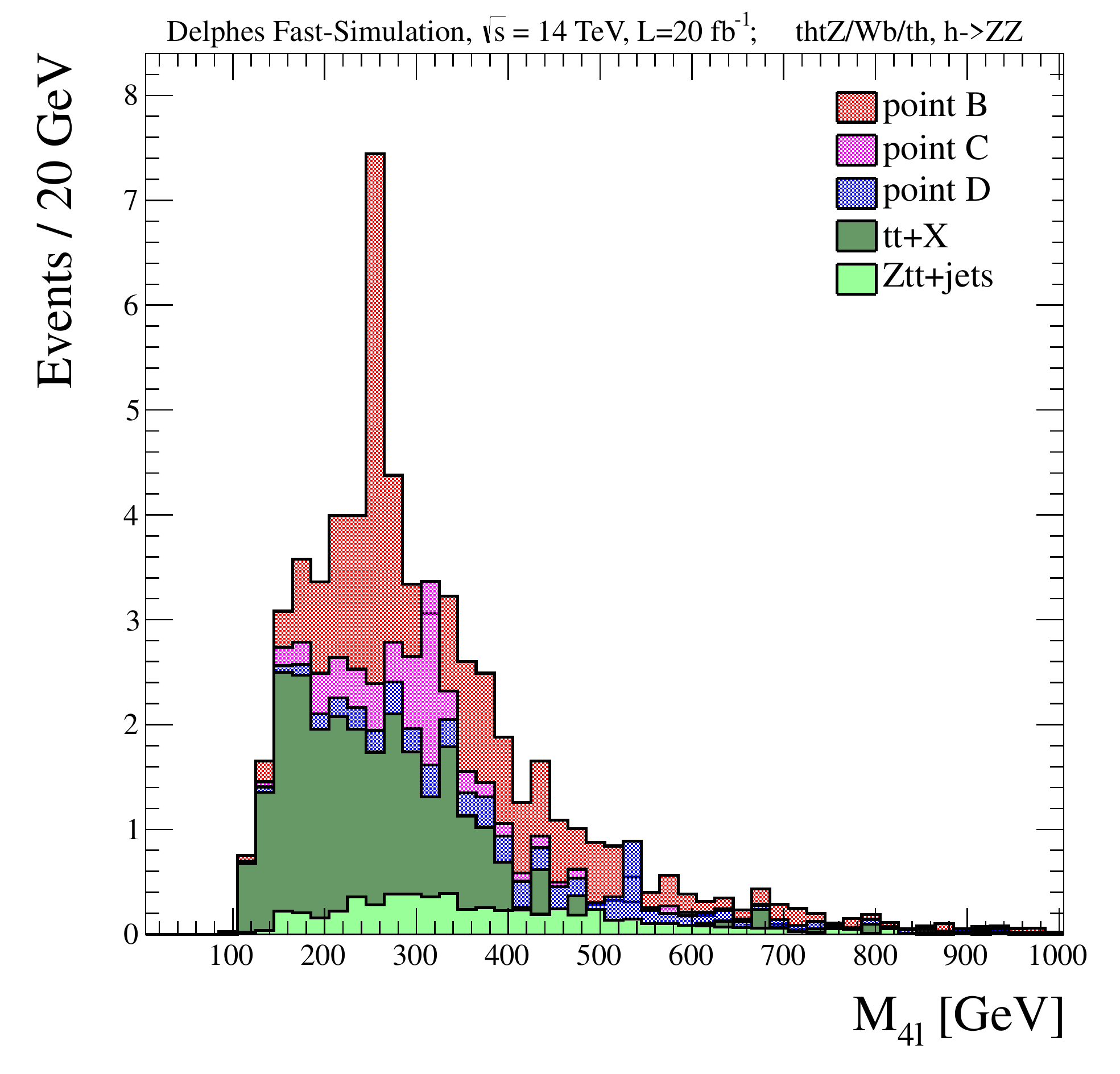}
&
\includegraphics[width=0.5\textwidth]{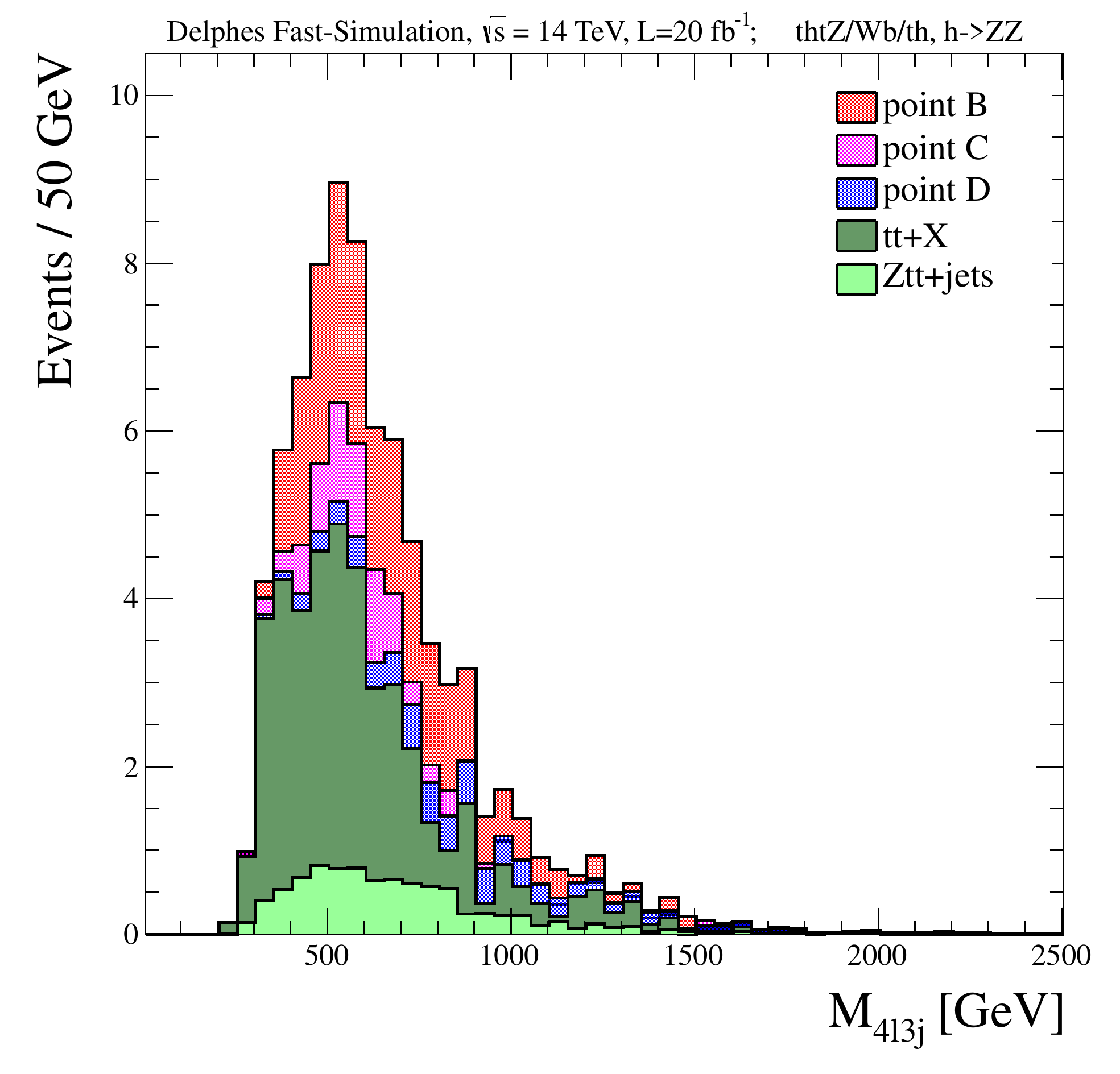}
\\
\end{tabular}
\caption{ Distributions of the  four lepton invariant mass, $M_{4l}$,  and of the  reconstructed heavy top-partner mass, $M_{4l3j}$ (see text), after all selection cuts. Signal and background distributions, normalized to the expected events for 20~fb$^{-1}$,  are shown stacked to indicate their relative contributions. Signal distributions for points B, C and D correspond to Higgs mass $m_h$ = 255, 320 and 540 GeV, respectively.}
\label{VTH:ZZmasses}
\end{center}
\end{figure}

The corresponding distributions of the invariant mass of the four leptons, $M_{4l}$, for signals and backgrounds are  shown in Figure~\ref{VTH:ZZmasses}. The mass peak corresponding to the generated Higgs mass is reconstructed at the correct mass value with good resolution ($\Delta M_{4l}/ M_{4l}\lesssim 10\%$).  Signal events, with reconstructed mass significantly lower, or higher, than the generated Higgs mass value, are in general events where one lepton from the $h\rightarrow ZZ\rightarrow 4 l$ decay  does not pass the lepton selection criteria,   while another lepton from an accompanying $W$ or $ Z$ decays is selected to reconstruct the four lepton invariant mass. The background, mainly from the $t\bar t$ + jets process,  is mostly concentrated in the  $M_{4l}$ region below 200 GeV. Applying a lower cut on $M_{4l}$
removes a large fraction of the background and improves the signal significance, as shown in Table~\ref{VTH:tabzzyields}.

An additional distribution of the variable, $M_{4l3j}$, is shown after all but the $M_{4l}$ cuts. This variable  can help estimating directly the mass of the heavy top-like quark. It  is defined as the invariant mass of the system obtained by 
associating to the four lepton momentum direction the three closest jets (expected to come from the decay of the accompanying top-quark,  $t\to bW$, $W\to 2 \ jets$),  choosing among all the jets in the event those with the smallest $\Delta R\equiv\sqrt{\Delta \eta^2+\Delta\phi^2}$ with respect to the four lepton momentum direction.
The distributions of $M_{4l3j}$ for signals and backgrounds are shown in Figure~\ref{VTH:ZZmasses}. A broad peak, whose width is dominated by the jet energy resolution and jet combinatorics,  is observed centered at about the generated $t_2$ mass value (the $t_2$ physical width is given in Table~\ref{VTH:DoubletTable}) for the points B and C having larger $t_2$ pair production rates than for the set D.
Therefore, this mass reconstruction method, together with the four lepton mass reconstruction,  will give an early indication that a heavy particle 
has been produced in the collision and decayed into a Higgs boson (identified by the four lepton mass peak) plus three jets.
Another mass reconstruction method, described in the next section, that exploits the property of the two heavy particles having equal mass, will give another characterization of the origin of the observed signal. Ultimately,  a kinematic fit technique (utilized, for instance,  in top
quark mass measurements, see e.g. \cite{Abulencia:2005aj}), to test on an event-by-event basis the hypothesis that the event results from the production of two heavy particles of equal mass (see e.g. \cite{CMS-PAS-EXO-11-051}) 
-- one of which decays into two particles with known decays and masses (three jets, with invariant 
mass around the top mass, and four leptons, with invariant mass around  the value measured from the four lepton mass peak, see Figure~\ref{VTH:ZZmasses})  
-- can be properly applied in real data events, when  the detailed parameterizations of the experimental resolutions of the physics objects entering the fit can be obtained from data.

The numbers of signal and background events expected after all cuts, and for $M_{4l}>$200, 300 and 500 GeV, are shown in the  three rightmost columns of Table~\ref{VTH:tabzzyields} together with the expected statistical background standard deviation $\Delta B$   for the three signal points investigated in this search channel.
We estimate that, for an integrated luminosity of 20~fb$^{-1}$, a signal corresponding to point B  is detectable with a significance $S/\Delta B \sim 6$ (after  $M_{4l}>$200 GeV cut). For this benchmark point the $B_{t_2\rightarrow th}$ branching fraction is $\sim 0.6$. Then, for  similar $t_2$ cross section, $t_2$ mass and  $h$ mass values, a signal could be discovered, with $S/\Delta B \gtrsim 5$, for
 $B_{t_2\rightarrow th}\gtrsim 0.5$. For point C, with  $B_{t_2\rightarrow th}\sim 0.1$  we obtain a signal significance $S/\Delta B \sim 2$.
A signal significance of five and above is then obtained for $B_{t_2\rightarrow th}\gtrsim 0.25$.
For smaller $B_{t_2\rightarrow th}$ values, a larger integrated luminosity, or a more optimal analysis, would be needed for 
a signal significance of five or above.
For point D, with  $B_{t_2\rightarrow th}\sim 0.4$,  we obtain a signal significance $S/\Delta B \sim 1.4$ thus  even for maximal $B_{t_2\rightarrow th}\sim 1$ a signal significance of $\sim 3.5$ is attainable. 
Higher integrated  luminosity, or higher background rejection, depending on the $B_{t_2\rightarrow th}$ value (for $B_{t_2\rightarrow th}\sim 0.4$, about a factor of ten larger luminosity or background rejection)  is needed for 
a five sigma signal significance corresponding to the D scenario, with a  $t_2$ mass of about 800 GeV and a pair production cross section 
of about  0.3 pb. 
%%%-----------------------------------------------------------------------------------------------------
%%%---------------------- 20120325  MF UPDATED
%%%-----------------------------------------------------------------------------------------------------
\subsection{Search for $\bar{t}_2t_2\rightarrow th+X$ signal in the diphoton plus multijets channel}
%%%-------------------------------------------------------------------------------------------------
To study the sensitivity of a search for $thtZ,\ thbW$ and $thth$ final states, 
when the Higgs boson  is relatively light, as for points A1 and  A2,  with
$m_{h}=125$~GeV, we exploit the $h\to
\gamma\gamma$ decay channel. As for a light Higgs boson in the SM, in spite of the relatively small expected B$(h\to\gamma\gamma)$ value, 
this is expected to be a channel with good signal sensitivity, due to the good diphoton mass resolution,
allowing to identify the Higgs mass signal over a background that can be well measured in the side-bands.

The background processes considered in this study are listed in Table~\ref{VTH:diphoyield}.
To reduce generation time for the background events,  we have applied cuts on {\small \tt ALPGEN} generated parton quantities, that are looser than the cuts applied on the reconstructed physics objects. The generator level cuts are: photon $p_T >20$ GeV and $|\eta|<$3.0, jet $p_T >20$ GeV and $|\eta|<2.5$, $\Delta R_{\gamma j}<0.4$,  $\Delta R_{\gamma l}<0.4$, , $\Delta R_{j j}<0.4$ and  $\Delta R_{t\bar{t}}<0.4$. Background cross sections for these generator level cuts as calculated by {\small \tt ALPGEN} are listed in Table~\ref{VTH:diphoyield}. 
An additional background is  the $\gamma\gamma$+jets process.  A relatively large inclusive diphoton cross section is expected~\cite{PhysRevLett.108.072001} and is going to be measured at the LHC, together with the contribution from fake diphoton pairs. 
The inclusive diphoton cross section in the diphoton mass region of interest can be large (order of pb), but we expect that the large jet multiplicity requirement scales down this contribution by at least three orders of magnitude, thus reducing its cross section to few fb. An additional handle to minimize this background  will be the b-tagging, that can reduce the contribution from light flavored multi-jets events by two or three order of  magnitudes (as discussed in the previous 
section), while retaining a large fraction of the signal events containing at least two  b-jets. However, b-tagging will not effectively reduce backgrounds with t-quark pairs in the final state, as in the case of  $t\bar{t} \gamma\gamma$+ jets process. Thus, conservatively we have not exploited b-tagging  for this analysis of  the $h\to\gamma\gamma$ channel.

Signals from   $thbW,\ thtZ,\ thth$, $h\to\gamma\gamma$, final states are characterized by a large number of energetic jets,
from top and heavy vector boson decays, in addition to the two high transverse momentum photons from the Higgs decay. 
The event selection consists of the following criteria:
\begin{itemize}
\item  %hoton identification and isolation requirements
%are applied that emulate the photon selection used in the CMS
%search for the SM Higgs boson to diphoton final states~\cite{Chatrchyan:2012tw}.
%% {CMS-PAS-HIG-11-030}. 
%%% change to the published paper
Photons are required to be within $|\eta|<2.5$ and isolated.
The isolation requirements imply that, within a cone $\Delta R=0.4$ around the photon direction, the charged particle energy measured in the tracker is $<2.0$ GeV, the electromagnetic calorimeter (ecal) energy in the cone is $<4.2$ GeV, the hadronic calorimeter (hcal) energy in the cone is  $<2.2$ GeV and the ratio between the ecal and hcal energy in the cone is $<0.05$. Two isolated photons are required, with the leading photon $p_T>45$ GeV and the second photon $p_T>30$ GeV. The invariant mass of the two photon is required to be $M_{2\gamma}>90$ GeV.
\item Hadronic jets are counted  if they have $p_T>$30 GeV and $|\eta|<2.4$. Events are required to have a number of jets $N_j>$ 6, at preselection, or $N_j>$ 8, for the final selection, after which the signal sensitivity is evaluated in a sliding window in the diphoton invariant mass $M_{2\gamma}$.  
\end{itemize}
%%%-----------------------------------
\begin{figure}[htbp]
\begin{center}
\begin{tabular}{cc}
\includegraphics[width=0.5\textwidth]{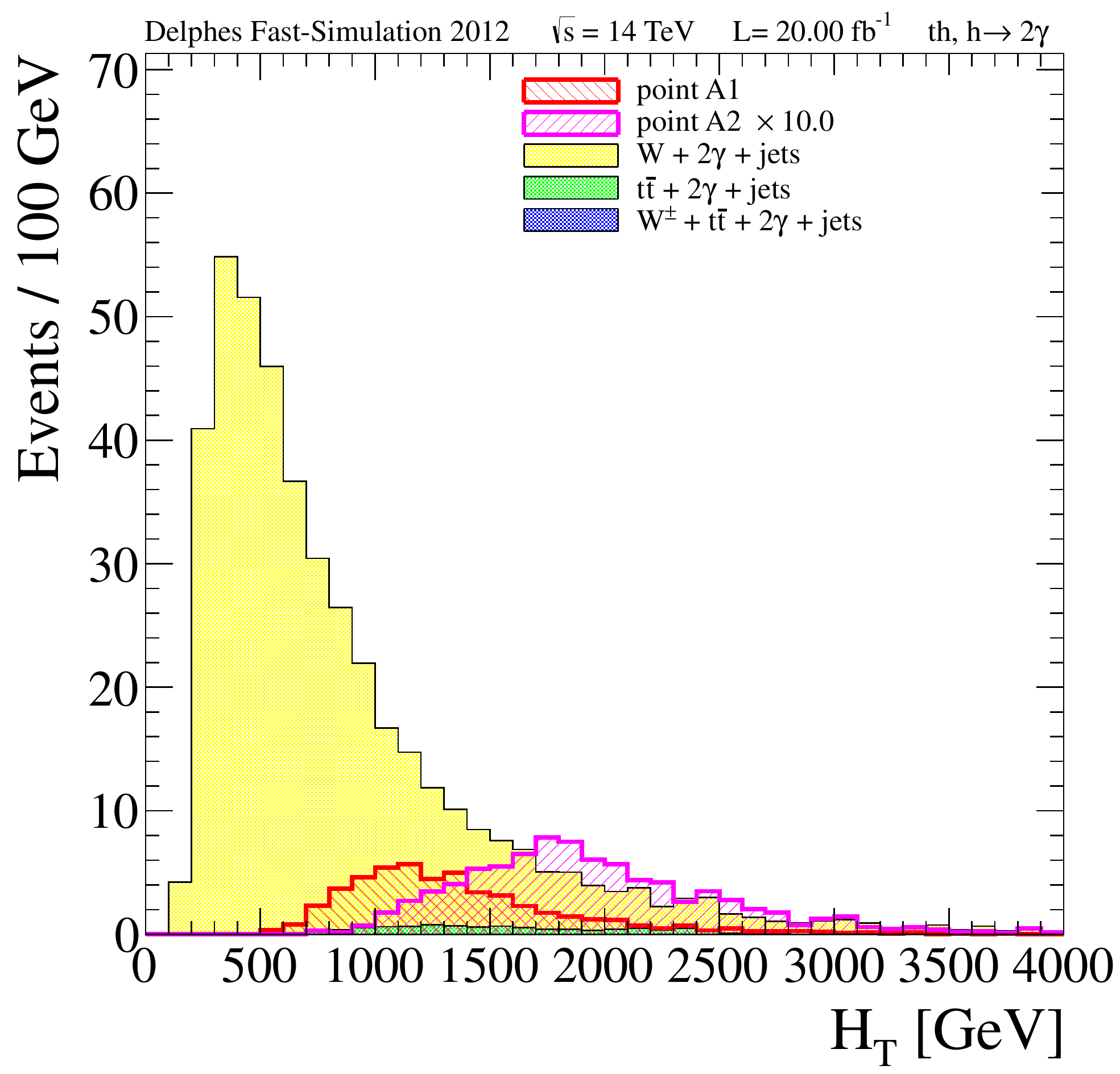}
&
\includegraphics[width=0.5\textwidth]{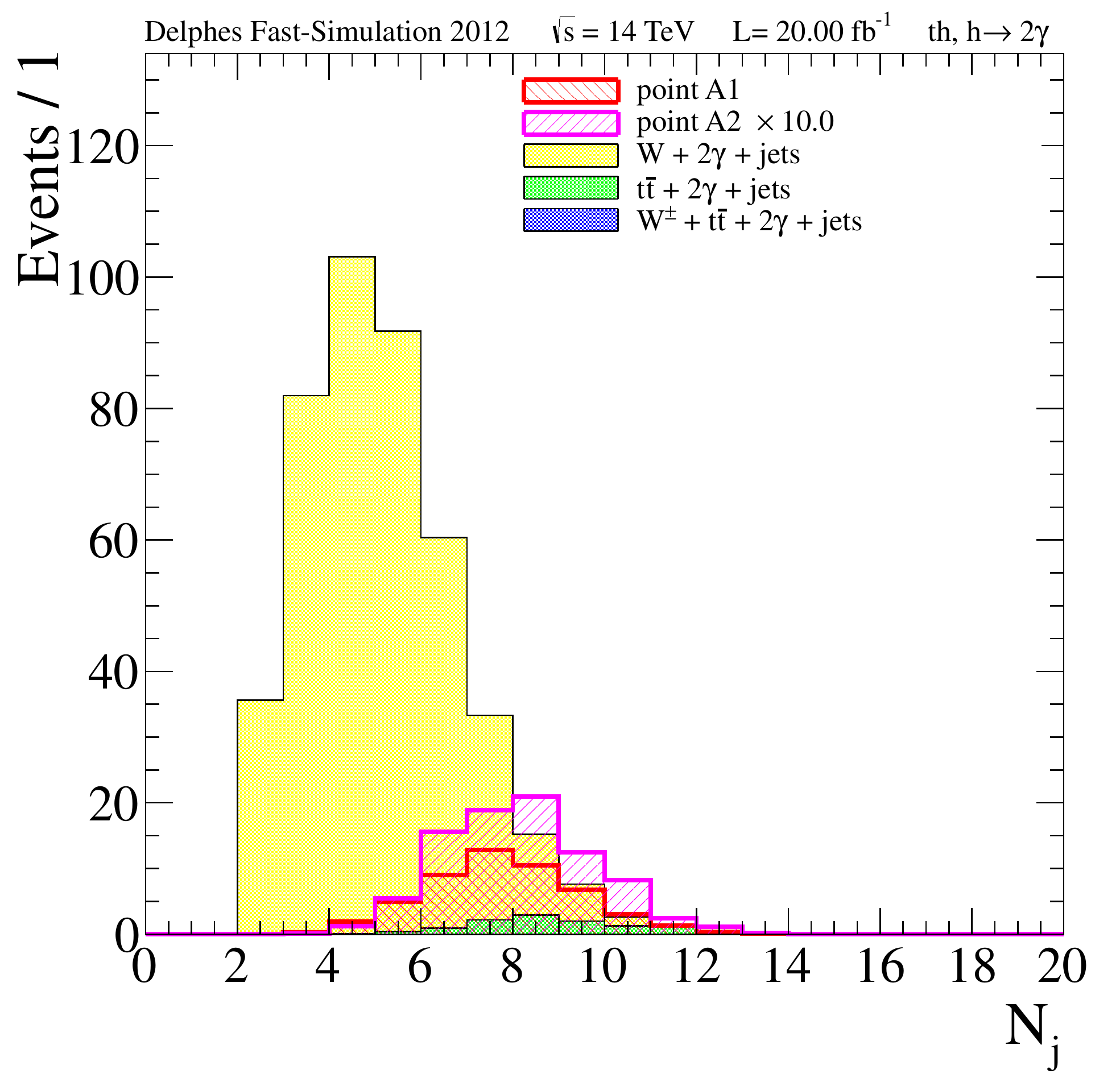}
\\
\end{tabular}
\caption{Distributions of the $H_T$ variable and the jet multiplicity $N_j$ after diphoton requirements, including a diphoton invariant mass cut $M_{2\gamma}>90$ GeV.}
\label{VTH:ggnjets}
\end{center}
\end{figure}
%%%-----------------------------------
Figure~\ref{VTH:ggnjets} shows the  $H_T$ and $N_j$ distributions after the 
after diphoton requirements, including a diphoton invariant mass cut $M_{2\gamma}>90$ GeV.
The distributions are shown for an integrated luminosity $L=20$ fb$^{-1}$, with the signal expectation for point A2 multiplied by a factor of ten, for shape comparison. Because of the relatively light $t_2$ mass in point A1, an $H_T>1000$ GeV cut, as applied for points B, C and D, would reject a significant fraction  of this signal. A lower $H_T$ cut would not improve significantly the signal-over-background ratio after the   $N_j$ requirement, thus no $H_T$ cut is applied in this channel. Expectations for the signals, by final state,  and the backgrounds processes, after the selection cuts, for an integrated luminosity of  $L=20$ fb$^{-1}$, are shown in Table~\ref{VTH:diphoyield}, where the following quantities are reported:
\begin{itemize}
\item $Y_S$, the signal event number produced,  per unit of integrated luminosity, given for each signal parameter set and final state under study,  as reported in Table~\ref{VTH:signalyield};
\item $\sigma$, the background cross sections, as calculated from the ALPGEN generator,  for the background processes included in the present study; 
\item $N_{2\gamma}$, the number of events after diphoton and  $M_{2\gamma}>90$ GeV requirements;
\item $N_{N_j>6}$ and $N_{N_j>8}$ the number of events after the additional jet multiplicity cuts;
\item $S$ and $B$, the total number of signal (summed over all signal final states) and background (summed over all background processes),
after a cut  $M_{2\gamma}$, with $\Delta B$ being the expected statistical uncertainty (standard deviation) on the total number of background events. An estimation of the expected signal sensitivity can then be evaluated as the ratio  $S/\Delta B$, measuring the signal in terms of background standard deviations.
\end{itemize}
%%%----------------------------------------------
\begin{table}[htbp]
\begin{center}
\begin{tabular}{|c|c|c|c|c|c|c|}
\hline
        & \multicolumn{4}{c|}{Signal: $thbW / thtZ / thth$, $h\to \gamma\gamma$} &  \multicolumn{2}{c|}{Total  signal $S$, $M_{2\gamma}$(GeV) cut }\\ \hline
 Parameter set & $Y_S$ (fb) &  $N_{2\gamma}$ & $N_{N_j>6}$ & $N_{N_j>8}$ &  \hspace{0.75cm}$M_{2\gamma}>90$\hspace{0.75cm} & $M_{2\gamma}\in $[115,135]  \\ \hline
 Point A1  &   3.1 / 10.1 / 22.1     &  5.4 /  15.4 / 30.4 & 4.1 /  12.7 /  27.1   & 1.4 / 6.2 / 14.5 & 22.1   & 17.7 \\   \hline 
 Point A2  &   0.16 / 0.73 / 4.1 &  0.34 / 1.35 / 6.98 & 0.27 / 1.17 / 6.6   & 0.10 / 0.62 / 3.8  & 4.5  &  3.2\\         
\hline
         &  \multicolumn{4}{c|} {Background }& \multicolumn{2}{c|}{Total background $B$, $M_{2\gamma}$(GeV) cut}\\ \hline
 Process  & $\sigma$(fb)          &   $N_{M_{2\gamma}>90}$ & $N_{N_j>6}$ & $N_{N_j>8}$ &  $M_{2\gamma}>90$ &$M_{2\gamma}\in $ [115,135] \\ \hline
 $W \gamma\gamma$ + jets             & 450             &  422        &  110       & 19.6    &    27.8     &  4.38  \\
\cline{6-7}
 $t\bar{t} \gamma\gamma$ + jets  & 15.5             & 11.8        &   11.3      &  8.18          &      \multicolumn{2}{c|}{Statistical $\Delta B$, $M_{2\gamma}$(GeV) cut} \\ \cline{6-7}
 $ t\bar{t}W\gamma\gamma$ + jets  &  0.0678 & 0.0577        & 0.0515 &  0.0272          &   $M_{2\gamma}>90$ & $M_{2\gamma}\in $[115,135]     \\ \cline{6-7}
% $\gamma\gamma$ + jets  &  reducible 
 & &        &         &            &   5.3    &  2.1 \\
\hline
\end{tabular}
\end{center}
\caption{Cross section and number of expected events after different selection cuts (see text), 
 for an integrated luminosity of 20~fb$^{-1}$ at $\sqrt{s}=14$ TeV, 
in the search for $thbW,$ $thtZ$ and  $thth$ final states, with $h\rightarrow \gamma\gamma$, in the diphoton plus multijets channel.  
The number of events is given separately for each signal final state ($thbW / thtZ / thth$) and for each background process. 
In the two  rightmost columns, 
$S$, the total number of signal events (summed over all signal final states),  and $B$, the total number of background events 
(summed over all background processes), are given after all selection cuts in column  ``$M_{2\gamma}>90$"  and
within a mass window $M_{2\gamma}\in $[115,135] GeV in the next column. 
The $S$ values for points A1 and A2 are given in the same lines as indicated in column ``Parameter set". 
The $B$ values in the column  ``$M_{2\gamma}>90$" are the sum of the individual background contributions as reported in the background column under  ``$N_{N_j>8}$".   
The quantity $\Delta B=\sqrt{B}$ is  the expected statistical uncertainty on the total number of background events. In the text, the signal sensitivity is given as the ratio  $S/\Delta B$.}
\label{VTH:diphoyield}
\end{table}

\begin{figure}[htbp]
\begin{center}
\begin{tabular}{cc}
\includegraphics[width=0.5\textwidth]{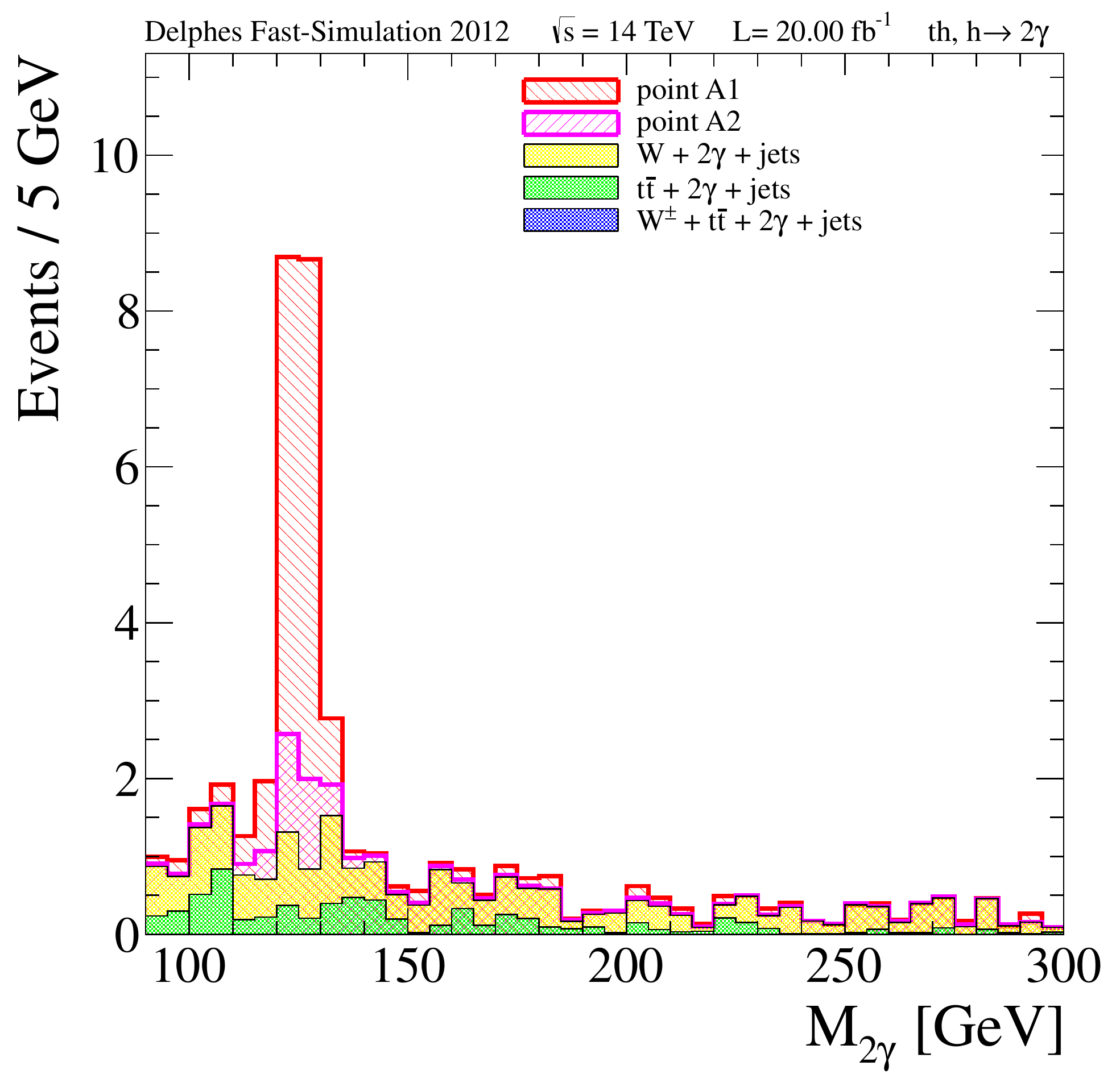}
&
\includegraphics[width=0.5\textwidth]{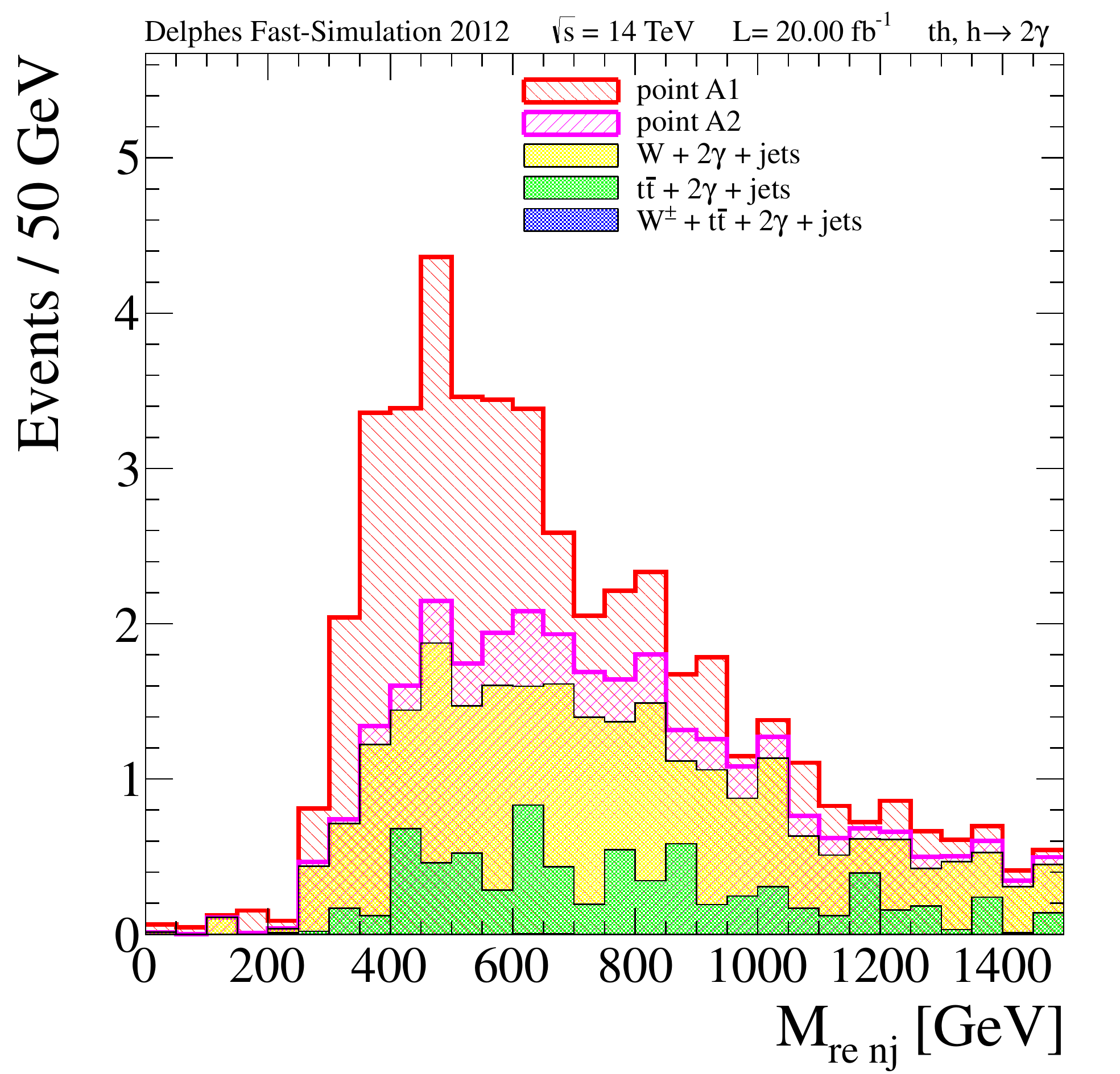}
\\
\end{tabular}
\caption{Distributions of the diphoton mass, $M_{2\gamma}$,  and of the reconstructed heavy recoil mass, $M_{re\ nj}$ (see text), after all selection cuts. Signal and background distributions, normalized to the expected events for 20~fb$^{-1}$,  are shown stacked to indicate their relative contributions. Signal distributions for points A1 and A2  correspond to a Higgs mass $m_h$ = 125  GeV.}
\label{VTH:ggmass}
\end{center}
\end{figure}

%---------------------------------------------------
In virtue of the good diphoton mass resolution ($\sim$$1\%$ to $5\%$ for a 110 to 130 GeV Higgs boson, depending on photon selection criteria) of the detector, a highly discriminating
quantity after selection is the reconstructed diphoton mass. The diphoton invariant mass distribution for signals and backgrounds events,
after the $N_j>8$ requirement,  is shown in Figure~\ref{VTH:ggmass}. Also shown is the distribution of an additional variable, $M_{re\ nj}$, that could help estimating the mass of the heavy top-like quark. This is defined as the invariant mass of the system of $n$ jets recoiling against the diphoton+$m$ jets system, where the $m$ jets are the closest in $\Delta R$ to the diphoton direction. The recoiling $n$ jets, with $n=N_j-m$, are counted to minimize, among all $m$ and $n$ 
choices in the event, the difference between the mass of the diphoton$+m$ jets system and the  mass of the recoiling $n$ jets. The goal of this procedure is to separate the event into two hemispheres, one identified by the diphoton+$m$ jets system and the other by the recoiling  $n$ jets, with $m$ and $n$ chosen in such a way that the invariant masses of the 
physics objects in the two hemispheres are about the same (within resolution), as in the case of two heavy particles decaying in diphoton plus jets, in one hemisphere,  and  into jets, in the other hemisphere. 
\\ On the distributions of $M_{re\ nj}$ for signals and backgrounds, after all but the $M_{2\gamma}$ window cut, shown in Figure~\ref{VTH:ggmass}, a broad peak, whose width is dominated by the jet energy resolution,  is observed  
centered at about the generated $t_2$ masses. Hence, this mass reconstruction method will provide an early indication that two heavy particles of equal mass have been produced in the event and a first estimation of the   $t_2$ mass. In real data, where the detailed measurements of the experimental resolutions are available, kinematic fits can be applied to test the pair production and decay hypothesis and possibly to reduce the uncertainty on the $t_2$ mass estimation.
%%%------------------------------------------
%%%------------------------------------------

The total number of signal and background events, after diphoton and jet multiplicity requirements, with a diphoton mass in the window between 115 GeV and 135 GeV, is also given in Table~\ref{VTH:diphoyield}. 
We estimate that for an integrated luminosity of 20~fb$^{-1}$, a signal, corresponding to the benchmark point A1, is
detectable with a significance $S/\Delta B\sim 9$. For this benchmark point  the $B_{t_2\rightarrow th}\sim 0.6$. Then, for a similar $t_2$ cross section and $t_2$, $h$ mass values, a signal could still be discovered, with a signal significance of about or above five, 
if  $B_{t_2\rightarrow th}\gtrsim 0.3$. For point A2, where the $\sigma_{\bar t_2t_2}$ is about an order of magnitude lower than in point A1, while $B_{t_2\rightarrow th}\sim 0.8$, the expected signal significance is $\sim 1.5$. Then, to reach a signal significance of five, at least a factor of ten larger luminosity, or a better background rejection, is needed to observe this signal.
%%%--------------------------------------------
\section{CONCLUSIONS}
In this paper, we have shown that there exist regions of the parameter space in beyond SM scenarios, allowed by present
phenomenological and experimental constraints, in particular by  the direct Higgs searches, where heavy vector-like
top partner production at the LHC may give rise to new Higgs production channels, over a large Higgs mass range, extending
from $\sim$~120 GeV to more than 500 GeV. Indeed the SM Higgs mass exclusion, in the range between $\sim130$ and 600 GeV, does not hold
in these beyond SM scenarios. 
In this context, we have also illustrated how the possible observation of a $125$~GeV Higgs signal in the diphoton channel at LHC, with a 
cross section larger than expected in the SM, could be compatible with the presence of vector-like heavy quarks.

We have shown that  for a variety of characteristic top-partner and Higgs mass values, the expected heavy top pair production cross sections and decay branching fractions lead to a signal of Higgs production (from the decay of the heavy 
top quark) which could be  detectable at  the 14 TeV LHC, with an integrated luminosity between $\sim$20~fb$^{-1}$ and $\sim$200~fb$^{-1}$,  over a large Higgs mass range ($\sim$120 GeV to $\sim$550 GeV). The good diphoton and four lepton invariant mass  resolution ($\lesssim 10\%$), in spite of the large number of jets present in the signal events, makes the discovery possible with moderate luminosity.  Finally, we have proposed new mass variables from which the heavy top mass could be estimated. 
%%%--------------------------------------------
%%%-------------------------------------------------------------------------------------------
%%%---------------- MF Revised 20120305 END -----------------------------------------------
%%%-------------------------------------------------------------------------------------------
\section*{ACKNOWLEDGEMENTS}
The work of G.M. is supported by the `Institut Universitaire de France' as well as the ANR {\it CPV-LFV-LHC} under project \textbf{NT09-508531} and {\it TAPDMS} under project \textbf{09-JCJC-0146}. 
D.K.G. acknowledges partial support from the Department of Science and Technology, India under the grant
SR/S2/HEP-12/2006.
%%%-----------------------------------------
\bibliography{sampleVTHiggs.12}

\end{document}